\documentclass[twocolumn]{aastex63}

\shorttitle{A Volatile-Poor Formation of LHS~3844b}
\shortauthors{Stephen R. Kane et al.}

\begin{document}

\title{A Volatile-Poor Formation of LHS~3844b based on its Lack of
  Significant Atmosphere}

\author[0000-0002-7084-0529]{Stephen R. Kane}
\affiliation{Department of Earth and Planetary Sciences, University of
  California, Riverside, CA 92521, USA}
\email{skane@ucr.edu}

\author[0000-0002-9288-3482]{Rachael M. Roettenbacher}
\affiliation{Yale Center for Astronomy \& Astrophysics, Yale
  University, New Haven, CT 06520, USA}

\author[0000-0001-8991-3110]{Cayman T. Unterborn}
\affiliation{School of Earth and Space Exploration, Arizona State
  University, Tempe, AZ 85287, USA}

\author[0000-0002-6943-3192]{Bradford J. Foley}
\affiliation{Department of Geosciences, Pennsylvania State University,
  University Park, PA 16802, USA}

\author[0000-0002-0139-4756]{Michelle L. Hill}
\affiliation{Department of Earth and Planetary Sciences, University of
  California, Riverside, CA 92521, USA}


\begin{abstract}

Exoplanet discoveries have reached into the realm of terrestrial
planets that are becoming the subject of atmospheric studies. One such
discovery is LHS~3844b, a 1.3 Earth radius planet in a 0.46 day orbit
around an M4.5-5 dwarf star. Follow-up observations indicate that the
planet is largely devoid of substantial atmosphere. This lack of
significant atmosphere places astrophysical and geophysical
constraints on LHS~3844b, primarily the degree of volatile outgassing
and the rate of atmosphere erosion. We estimate the age of the host
star as $7.8\pm1.6$~Gyrs and find evidence of an active past
comparable to that of Proxima Centauri. We use geodynamical models of
volcanic outgassing and atmospheric erosion to show that the apparent
lack of atmosphere is consistent with a volatile-poor mantle for
LHS~3844b. We show the core is unlikely to host enough C to produce a
sufficiently volatile-poor mantle, unless the bulk planet is
volatile-poor relative to Earth. While we cannot rule out a giant
impact stripping LHS 3844b's atmosphere, we show that this mechanism
would require significant mantle stripping, potentially leaving
LHS~3844b as an Fe-rich "super-Mercury". Atmospheric erosion by
smaller impacts is possible, but only if the planet has already begun
degassing and is bombarded by $10^3$ impactors of radius 500--1000~km
traveling at escape velocity. We discuss formation and migration
scenarios that could account for a volatile-poor origin, including the
potential for an unobserved massive companion planet. A relatively
volatile-poor composition of LHS~3844b suggests that the planet formed
interior to the system snow-line.

\end{abstract}

\keywords{planets and satellites -- interiors planetary systems --
  techniques: photometric -- stars: individual (LHS~3844)}


\section{Introduction}
\label{intro}

Exoplanetary science has advanced to a regime where terrestrial
planets are routinely discovered orbiting other stars. For example,
the {\it Kepler} mission discovered several hundred terrestrial
planets during the primary four years of observations
\citep{borucki2016}, of which tens were found to lie within their
star's Habitable Zone \citep{kane2016c}. Similar terrestrial planet
discoveries are now continuing through the use of the Transiting
Exoplanet Survey Satellite ({\it TESS}), launched into Earth orbit in
2018 \citep{ricker2015}. It is expected that the relative brightness
of {\it TESS} host stars will allow the discovered planets to become
prime targets for atmospheric characterization using follow-up
facilities, such as the James Webb Space Telescope ({\it JWST})
\citep{kempton2018,ostberg2019}. Such observations will allow tests of
composition and atmospheric erosion scenarios
\citep{owen2019a,rodriguezmozos2019}. Such erosion scenarios are
considered particularly relevant to low-mass stars that can have
extended periods of high activity, such as the case of TRAPPIST-1
\citep{roettenbacher2017,dong2018a}.

One of the very early discoveries announced using data from the {\it
  TESS} mission was the detection of a planet orbiting the star
LHS~3844 \citep{vanderspek2019} with an orbital period of $\sim$11
hours. The planet is likely terrestrial, with a radius of $R_p = 1.303
\pm 0.022$~$R_\oplus$, although no strong constraints on the planet
mass have yet been established. The importance of the planet was
raised significantly by \citet{kreidberg2019b}, who reported
observations of the LHS~3844 system using the {\it Spitzer} space
telescope. Their analysis of the {\it Spitzer} data indicated that the
planet does not have a thick atmosphere, down to a limit of 10~bar,
consistent with severe atmospheric erosion of the primary
atmosphere. It remains to be seen whether similar atmospheric erosion
effects are common among terrestrial planets around low-mass stars,
and how these effects depend on planetary mass and age.

The buildup of a secondary atmosphere is a coupled astronomical and
geological process. Geologically, melting in the shallow subsurface of
the planet releases volatile gasses to the atmosphere
\citep[e.g.,][]{holland1984,zhang2014,foley2016,noack2017a,tosi2017},
while atmospheric erosion due to stellar activity removes this
atmosphere over time \citep{sakuraba2019}. The special case of
LHS~3844b currently having little to no atmosphere then allows us to
simultaneously constrain both the astrophysical environment the planet
has experienced as well as aspects of its geological evolution and
composition.

Here, we present the results of a combined stellar and geophysical
study that aims to provide context for the observed lack of thick
atmosphere for LHS~3844b. In Section~\ref{host} we provide estimates
for the age and activity of the host star based on known stellar
parameters. In Section~\ref{planet} we calculate a geologically
motivated planetary mass and density, and place the incident flux
received by the planet in the context of other similar
exoplanets. Section~\ref{atmosphere} provides a detailed description
of our interior model and method for estimating potential degassing
time scales and atmospheric erosion scenarios. The results of our
model calculations, including the time evolution of atmospheric
pressure, are provided in Section~\ref{results}. In
Section~\ref{discussion}, we discuss the implications of the
geodynamical models, core and mantle volatile inventory, and stellar
erosion factors for the formation and migration scenarios of the
planet relative to the snow-line. We also include a detailed
description of the effect of potential small and giant impactors on
the atmospheric evolution and evaluate the likelihood of impacts as
the source of the observed constraints for the LHS~3844b
atmosphere. We provide concluding remarks in
Section~\ref{conclusions}, including future implications and
applications of this work.


\section{Host Star Properties}
\label{host}

We adopt the accumulated stellar parameters provided by
\citet{vanderspek2019}, which include mass, radius, and effective
temperature of $M_\star = 0.151 \pm 0.014$~$M_\odot$, $R_\star = 0.189
\pm 0.006$~$R_\odot$, and $T_\mathrm{eff} = 3036 \pm 77$~K
respectively. \citet{vanderspek2019} further provide a spectral type
of M4.5 or M5, and estimate a stellar rotation period of $128 \pm
24$~days based on time series photometry. For this study, we further
estimate the age and the activity properties of the star.


\subsection{Age}
\label{age}

Gyrochronology is a useful tool for constraining stellar ages by
combining observed rotation periods with other intrinsic stellar
properties \citep{barnes2007f,angus2019a}. However, the methodology
becomes increasingly uncertain for very low-mass stars where rotation
rates are poorly calibrated with age estimates
\citep{angus2019b,gallet2019a}. In order to use gyrochronology, stars
are required to spin down over time due to angular momentum loss.
This angular momentum loss has been shown to be strongly tied to the
stellar open flux, which is the magnetic flux found in the open field
lines of the stellar magnetic field that carries winds away from the
star \citep{mestel1987a}. For stars with convective outer envelopes,
this open flux can be estimated to be a dipolar field
\citep[e.g.,][]{petit2008}. However, low-mass M dwarfs that are fully
convective can have their open flux estimated as either a dipolar
field or a more complex field structure
\citep[e.g.,][]{donati2008c}. Even knowing the structure of the
magnetic field does not allow observers to understand the nature of
the spin-down of these low-mass stars \citep{see2017}.

\citet{newton2016a} used photometry from the MEarth Project to obtain
rotation periods and combined that information with a proper motion
survey \citep{lepine2005a} to estimate the ages of low-mass stars
based on how their rotation period correlates to where in the galaxy
the star is found. LHS~3844 has a reported rotation period of $128 \pm
24$ days \citep{vanderspek2019} from ground-based MEarth monitoring,
but was not included in the \citet{newton2016a} sample. For stars with
rotation periods greater than 70 days, \citet{newton2016a} estimated
that stars have an average age of $5.1^{+4.2}_{-2.6}$ Gyr.  Their
sample included 28 stars with an average rotation period of 106.2
days. \citet{engle2018a} provided a rotation-age relationship for M
dwarfs, making use of ten years of ground-based photometry and the
stars' membership or probable membership in a cluster or group.  The
relationship for M dwarfs with spectral types M2.5-M6, within which
LHS~3844 falls, suggests an age of $7.8 \pm 1.6$ Gyr, which is
consistent with the \citet{newton2016a} estimate and provides the best
age estimate presently available for a late-M field star like
LHS~3844.


\subsection{Activity}
\label{activity}

\begin{figure}
  \includegraphics[angle=90,width=8.5cm]{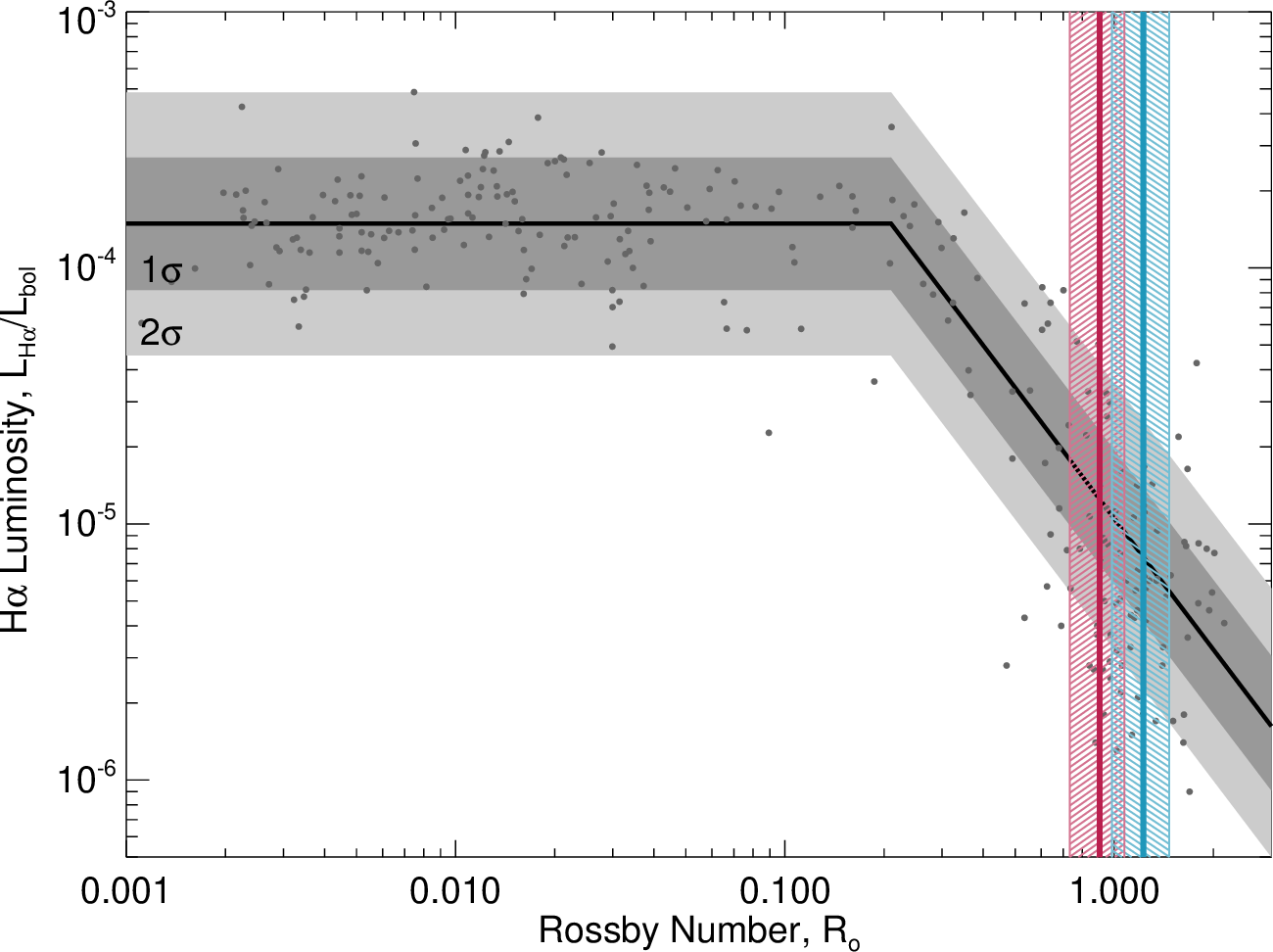}
  \caption{Activity-rotation relationship diagram with H$\alpha$
    luminosity, $L_{\mathrm{H}\alpha}/L_\mathrm{bol}$, plotted against
    Rossby number, $R_o$. The solid black line shows the
    activity-rotation relationship found by \citet{newton2017}.  The
    dark and light gray regions are their $1-$ and $2-\sigma$ errors,
    respectively, with the data points used for their calculations as
    dark grey dots. The pink hatched region (hatching lines increasing
    from left to right) indicates the range of $R_o$ estimated with
    photometric magnitudes, $V = 15.26 \pm 0.03$
    \citep{vanderspek2019} and $K_s = 9.145 \pm 0.023$
    \citep{skrutskie2006} using Equation 10 from
    \citet{wright2011d}. The teal hatched region (hatching lines
    decreasing from left to right) indicates the range of $R_o$
    estimated with the stellar mass, $M = 0.151 \pm 0.014 \ M_\odot$
    \citep[Equation 11 from][]{wright2011d}.  There are no recorded
    values of $L_{\mathrm{H}\alpha}/L_\mathrm{bol}$ for LHS~3844, but
    given the star's $R_o$, we suggest the star is likely only weakly
    active.}
  \label{fig:activityrotation}
\end{figure}

We calculated the Rossby number, or the ratio of stellar rotation
period to the convective turnover time, $R_o \equiv
P_\mathrm{rot}/\tau_\mathrm{conv}$, for LHS~3844 using Equations 10
and 11 from \citet{wright2011d}.  Equation 10 depends only upon $V -
K_s$ color and Equation 11 depends only upon stellar mass (in terms of
solar masses). For these quantities, we used parameters reported by
\citet{vanderspek2019}. We found that the Rossby number of LHS~3844 is
approximately $R_o = 0.90 \pm 0.18$, when using Equation 10 and the
appropriate photometric magnitudes, and $R_o = 1.22 \pm 0.25$, when
using Equation 11 and the mass of the star (see
Figure~\ref{fig:activityrotation}). Both of these values suggest that
the star presently is not strongly active. A Rossby number of $R_o \le
0.1$ indicates that a star is fully saturated in activity
\citep[e.g.,][]{wright2011d,wright2016c,newton2017,wright2018b}. That
is, faster rotation cannot increase the amount of activity observed
from the star. However, as Rossby numbers increase above $R_o = 0.1$,
due to increasing rotation periods, the amount of activity observed on
the star will decrease.  Largely due to its slow rotation period of
$P_\mathrm{rot} = 128 \pm 24$ days, LHS~3844 has a Rossby number
significantly higher than 0.1, and the range of values calculated
suggest that LHS~3844 is currently (weakly) active but not
saturated. The light curve of LHS~3844 shows evidence of rotational
modulation, likely due to starspots rotating in and out of view, but
no evidence of other activity events, such as stellar flares.

Because the nature of angular momentum loss for fully convective stars
is not well-determined, we are only able to suggest that, if LHS~3844
has spun down over time, then it was previously more active
\citep[e.g.,][]{reiners2008a}. While there is no evidence of flares in
the {\it TESS} light curve, if we assume that fully convective stars
spin down with the relationship given by \citet{engle2018a}, we can
use the activity of Proxima Centauri (hereafter Proxima Cen) to
understand how LHS~3844 may have behaved in the past. Proxima Cen is
an M5.5 dwarf with an age of approximately 6 Gyr and a rotation period
$P_\mathrm{rot} \approx 83$ days. \citet{vida2019b} found that Proxima
Cen had 1.49 flares/day in two sectors of {\it TESS} data (Sectors 11
and 12) with flares ranging from $10^{30}-10^{32}$~erg. They predict
that larger flares on the order of $10^{34}$~erg once every two years
for a star this active. While it is difficult to project stellar
activity forward or backward with fully convective M dwarfs, we can
use Proxima Cen as a proxy for how LHS~3844 could have behaved in the
past. We therefore adopt the range of atmospheric loss rates predicted
by \citet{kreidberg2019b}, who scaled the loss rate from Proxima Cen~b
to the value of 30--300~kg/s \citep{dong2017a}.


\section{Planetary Properties}
\label{planet}


\subsection{Mass, Radius, and Density}
\label{bulk}

In order to estimate volatile degassing rates for the planet, it is
necessary to determine fundamental planetary properties, including the
mean (bulk) density. The radius of the planet, provided by
\citet{vanderspek2019}, is $1.303 \pm 0.022$~$R_\oplus$. We utilize
the mass-radius-composition methodology of \citet{unterborn2019} in
order to estimate the mass of LHS~3844b. Knowing the planet lacks any
significant volatile atmosphere, we assume the planet is made entirely
of an FeO-free silicate mantle and pure liquid-Fe core. Mantle phase
equilibria and core radius fractions (CRF) are calculated using the
mass-radius-composition solver, ExoPlex \citep{unterborn2019}. We
adopt the thermodynamic equation of state data of \citet{stixrude2011}
to calculate mantle phase equilibria and use the fourth-order
Birch-Murgnahan equation of state for liquid Fe of
\citet{anderson1994}. These models accurately reproduce the Earth
\citep{unterborn2016} and provide more robust estimates for masses of
individual planets than do empirical models \citep[e.g.,
][]{zeng2016a}, and are also wholly self-consistent with mineral
physics experimental data \citep{unterborn2019}.

There are no estimates of the bulk composition of LHS~3844b, whether
directly through measurement of its density or indirectly from
estimates of host-star composition. As the mass of a rocky planet (for
a given radius) is most sensitive to the relative size of the central
Fe core, as defined by its core mass fraction (CMF), we choose two
end-member compositions of the modeled planet's bulk Fe/Mg of 0.6 and
1.5. This equates to CMFs of 0.25 and 0.45, respectively (n.b., the
Earth's CMF is 0.33). These end members encompass 80\% of all measured
stellar Fe/Mg within the Hypatia catalog
\citep{hinkel2014,unterborn2019}.

From this modeling, we estimate the mass of LHS~3844b to be 2.4
M$_\oplus$ for the small core case, 2.5 M$_\oplus$ for the Earth-like
core case, and 2.9 M$_\oplus$ for the large core case. These equate to
core radius fractions of 0.46, 0.55, and 0.59, respectively. We
calculate the bulk planet density then varying between 6 and 7.2
g~cm$^{-3}$ for the small and large core cases, respectively, making
them slightly more dense than the Earth (5.5 g~cm$^{-3}$). We note
that the probabilistic model forecasting tool provided by
\citet{chen2017} estimates a mass for the planet of
$2.20^{+1.57}_{-0.65}$~$M_\oplus$, leading to a bulk density of
$5.5^{+4.4}_{-1.7}$~g~cm$^{-3}$. While this result is consistent with
our model, we attribute the large uncertainties in the \citet{chen2017}
model result being due to LHS~3844b lying slightly above the boundary
between their defined ``Terran'' and ``Mini-Neptune'' mass
boundaries. Compared to the Terran planets, there is a larger
uncertainty in mass for a given radius within the mini-Neptune regime,
explaining the nearly factor of two uncertainty in the
\citet{chen2017} modeled bulk density.


\subsection{Incident Flux}
\label{flux}

\begin{figure}
  \includegraphics[angle=270,width=8.5cm]{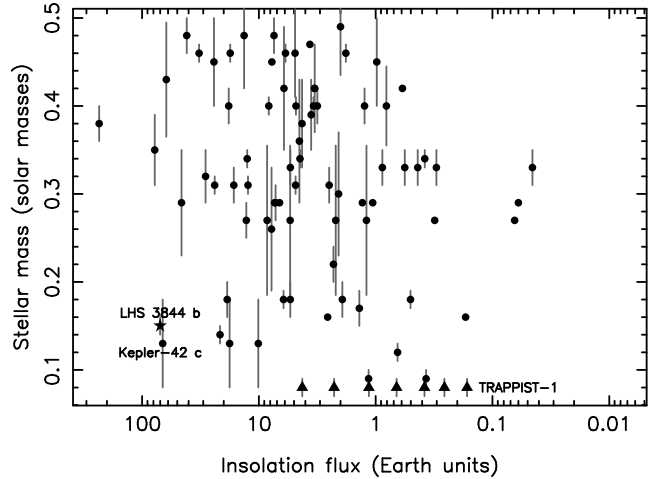}
  \caption{Plot of the host star mass and calculated insolation flux
    for all known exoplanets, where host star mass was restricted to
    $< 0.5$~$M_\odot$ and planet mass and radius where restricted to
    $< 10$~$M_\oplus$ and $< 2$~$R_\oplus$ respectively. The location
    of LHS~3844b is indicated by a star near the bottom-left corner of
    the plot. The TRAPPIST-1 planets are indicated by triangles.}
  \label{fig:flux}
\end{figure}

The incident (insolation) flux environment of the planet is a key
factor in the atmospheric evolution of the planet. To place LHS~3844b
in context, we extracted data for the confirmed exoplanets from the
NASA Exoplanet Archive \citep{akeson2013}. The data are current as of
2020 February 16. We selected those systems whose host star mass is
less than 0.5~$M_\odot$ and planets that either have a mass less than
10~$M_\oplus$ and/or a radius less than 2~$R_\oplus$. We used the
stellar information provided, specifically the stellar luminosities
and semi-major axes, to calculate the insolation flux for each
planet. These data are shown in Figure~\ref{fig:flux} for all of the
planets that met the above criteria. We do not include insolation
uncertainties in the plot because, although stellar mass uncertainties
are readily available, uncertainties on insolation and luminosities
are relatively scarce, resulting in a significant diversity of
insolation uncertainties (3--30\%) that would appear potentially
misleading. LHS~3844b receives $\sim$70 times the flux received by
Earth and is indicated by a star in the figure. The lowest mass host
star represented in the figure is the well-known TRAPPIST-1 system
\citep{gillon2017a}, where the planets are shown as triangles. The
closer a planet is located towards the bottom-left of the diagram, the
greater the risk of atmospheric erosion due to the combination of
stellar activity (including flare events) and the extreme flux
environment of the planet in proximity to the host star. LHS~3844b is
therefore in the highest atmospheric-loss risk regime compared with
all the other known planets. The only exception to that is
Kepler-42~c, a 0.73~$R_\oplus$ planet, which receives a flux $\sim$67
times the solar constant and orbits a star slightly less massive than
LHS~3844. Thus, Kepler-42~c is also highly likely to have experienced
significant atmospheric erosion, possibly to the point of complete
atmospheric desiccation if the planet has a low volatile inventory.


\section{Atmospheric Evolution}
\label{atmosphere}


\subsection{Atmosphere and Degassing}
\label{degassing}

\begin{figure}
  \includegraphics[width=8.5cm]{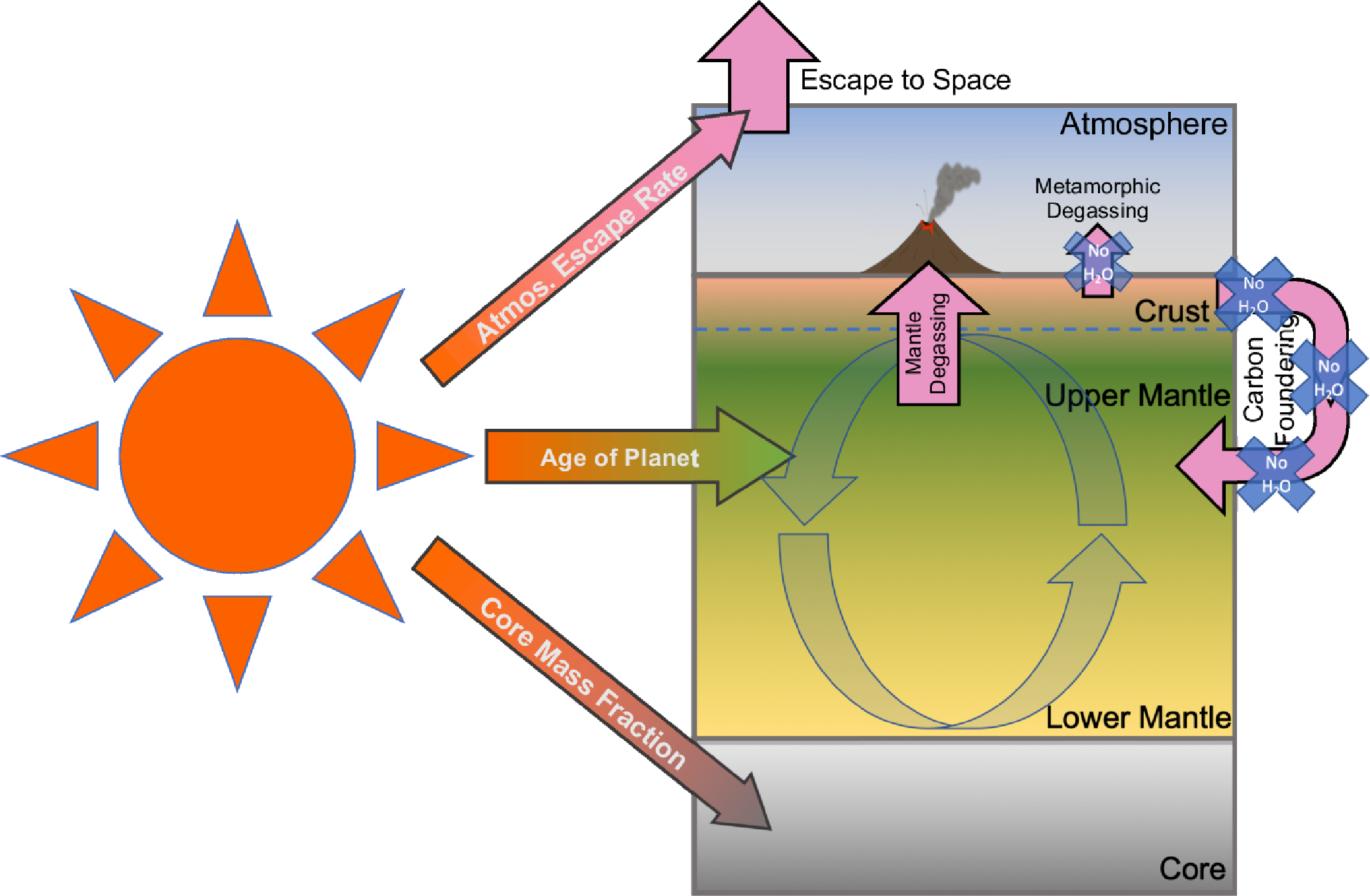}
  \caption{Schematic of our planetary evolution model of LHS-3844b
    undergoing stagnant lid tectonics adapted from
    \citet{foley2018a}. The host star provides an age of the system
    over which to model the geodynamic evolution (Section~\ref{age}),
    information on the atmospheric stripping rate through estimates of
    the XUV flux given its stellar type (Section~\ref{activity}), and
    the maximum core mass fraction from the measurements of the host
    star's Fe/Mg ratio, assuming all the Fe is in the core, which sets
    the size of the convecting mantle (Section~\ref{bulk}). Pink
    arrows represent the flux of volatile species into and out of the
    planet's atmosphere. We assume there was never water on LHS-3844b,
    as the presence of water is necessary for sequestering CO$_2$ into
    the mantle via crustal foundering, thus preserving it from
    atmospheric escape. This lack of water also prevents degassing
    from the crust via metamorphic reactions. Thus, the two primary
    fluxes for CO$_2$ in our model for LHS-3844b's evolution are the
    atmospheric escape rate (30--300~kg/s) and the rate of mantle
    degassing.}
  \label{fig:flowchart}
\end{figure}

After a planet has lost its primary H$_2$/He atmosphere, the creation
of a secondary atmosphere is primarily due to degassing of volatiles
from the interior by mantle volcanism. The rate of volatile outgassing
from the interior via mantle volcanism, and hence the total size of
atmosphere that can accumulate over time, is controlled by the thermal
evolution of the mantle
\citep[e.g.][]{tajika1992,hauck2002,grott2011b,tosi2017,dorn2018b,foley2018a}.
We use simple thermal evolution models to constrain outgassing history
and the resulting atmospheric mass, based on the works of
\citet{foley2018a} and \citet{foley2019}. Figure~\ref{fig:flowchart}
shows a schematic of our model including those values obtained by
analysis of LHS~3844 and the fluxes of CO$_2$ between the mantle,
crust, and atmosphere of LHS~3844b.

We seek to determine the combinations of planetary properties that can
lead to an atmosphere size $\leq 10$ bar, as observed for LHS~3844b,
by it's present day age. We therefore run a large suite of models
sampling from distributions of the key parameters that govern thermal
evolution and resulting atmospheric mass: the planetary volatile
budget, the atmospheric loss rate, the reference viscosity of the
mantle, the radioactive heating budget of the mantle, and the initial
mantle temperature. We track which models do or do not successfully
result in atmospheres $< 10$ bar in size. As the uncertainty range on
the age of LHS~3844 is 6.2--9.4~Gyrs, we determine for each model
whether it meets the constraints on atmosphere size by both the upper
and lower age bounds. The models assume LHS~3844b lies in a stagnant
lid regime of tectonics. While the tectonic mode of any exoplanet is
unconstrained, we assume a stagnant lid because the high surface
temperature of LHS~3844b and lack of surface water are expected to
disfavor plate tectonics
\citep[e.g.,][]{lenardic1994a,regenauerlieb2001,lenardic2008,landuyt2009a,korenaga2010b,foley2012}. As
a result of stagnant lid tectonics and the lack of surface water,
volcanism is expected to produce a basaltic crust covering the
planet's surface; this crustal composition is consistent with that
inferred by \citet{kreidberg2019b}.

The primary volatile species outgassed by volcanism on Earth are
H$_2$O and CO$_2$. As LHS~3844b receives too much radiation from the
star for liquid water to be stable at the surface \citep{tian2015a},
we focus on CO$_2$. Whether oxidized species, like CO$_2$, or more
reduced species, like CO or CH$_4$, are outgassed depends on the
oxidation state of the mantle
\citep[e.g.,][]{kasting1993e,gaillard2009}. Oxidation of Earth's
mantle is thought to occur just after planet formation by
disproportionation in the lower mantle
\citep[e.g.,][]{wade2005,wood2006}. This process is expected to occur
on rocky planets of Earth size or larger, as long as the mantle
mineral makeup is dominated by Mg-silicates. Assuming an oxidized
mantle for LHS~3844b is therefore reasonable. However, even if the
mantle of LHS~3844b is more reduced than Earth's, the same process of
mantle thermal evolution and degassing that we model still determines
how atmospheric mass evolves over time.

We modify the model of \cite{foley2018a} to apply to LHS~3844b. As the
exact mass and interior structure of LHS~3844b is unknown, we consider
three end members as outlined in Section~\ref{bulk}: a small core size
(46\% by radius), large core size (59\% by radius), and an Earth-like
core size (55\% by radius). For all cases, the planet radius is held
fixed at $1.3$~$R_\oplus$, as radius is tightly constrained, yielding
masses between 2.4 and 2.9 $M_\oplus$ for the small and large core
sizes, respectively. For the small core models, ExoPlex calculated the
average mantle density $\rho = 5400$~kg~m$^{-3}$ and gravity
$g=14.3$~m~s$^{-2}$; for the large core models $\rho =
5360$~kg~m$^{-3}$ \& $g=18.9$~m~s$^{-2}$; and for the Earth-like core
$\rho = 5380$~kg~m$^{-3}$ \& $g=15.7$ m~s$^{-2}$. As the planet cannot
sustain liquid water oceans or weathering, all CO$_2$ outgassed from
the mantle is assumed to accumulate in the atmosphere; this means that
there is no degassing flux from decarbonation of the crust, as there
is no weathering to deposit significant stores of carbon into the
crust. We also conservatively assume that melt becomes denser than the
solid mantle above a pressure of 6~GPa, and thus only melt formed at
pressures lower than this contributes to mantle outgassing. This
chosen melt density crossover pressure is on the low end of
experimental estimates \citep[e.g.,][]{suzuki2003,reese2007}, meaning
that uncertainty in the melt density will only serve to increase
outgassing rates and hasten the rapid mantle outgassing our models
predict (see Section~\ref{results}).

We assume the mantle has a finite CO$_2$ budget, which is set by an
initial concentration of CO$_2$ in the mantle, $C_\mathrm{conc}$. We
assume a range of $C_\mathrm{conc} = 10^{-4}-10^{-2}$ wt\%, varying
from the Earth's assumed CO$_2$ concentration of $\sim 10^{-2}$ wt\%
\citep[based on an estimate of $10^{22}$ mol of CO$_2$ in the mantle
  and surface reservoirs of Earth from][]{sleep2001b} to two orders of
magnitude lower. This range of CO$_2$ concentrations results in a
range of total mantle CO$_2$ budgets, $C_\mathrm{tot}$, of
$C_\mathrm{tot} \approx 2.5 \times 10^{20} - 2.5 \times 10^{22}$ mol
in our models. Note that a given mantle CO$_2$ concentration results
in slightly different total CO$_2$ budgets for the three end member
planet interior structures we model, as changing the core size changes
the total mass of the mantle. As the results show
(Section~\ref{results}), our chosen range of CO$_2$ budgets brackets
the key transition in model behavior, where models with low CO$_2$
budgets always end with a $<$ 10 bar atmosphere, while those with high
CO$_2$ budgets nearly always end with atmospheres $>$ 10 bar. We start
our models with all of the planet's CO$_2$ initially residing in the
mantle, and then we calculate how CO$_2$ accumulates in the atmosphere
over time by degassing. Some CO$_2$ may be outgassed during a putative
magma ocean stage, which is neglected here. Including magma ocean
degassing would only serve to hasten the rapid accumulation of mantle
C into the atmosphere that the large majority of our models predict.

\begin{figure*}
  \begin{center}
    \includegraphics[angle=270,width=16cm]{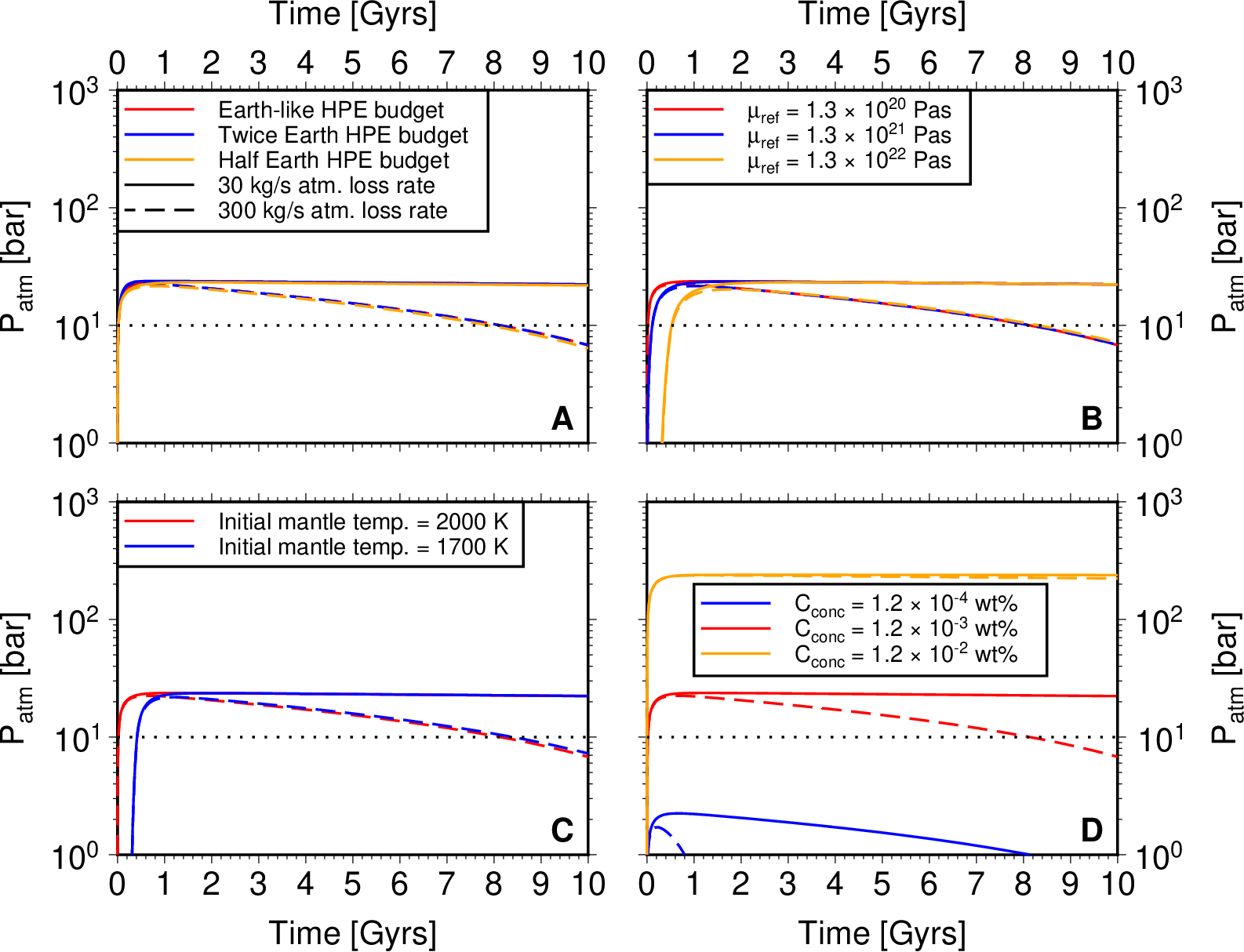}
  \end{center}
  \caption{Time evolution of total atmospheric pressure,
    $P_\mathrm{atm}$, for example models with an Earth-like core
    size. Variations from a baseline model with an Earth-like heat
    producing element (HPE) budget, mantle CO$_2$ concentration of
    $C_\mathrm{conc} = 1.2 \times 10^{-3}$ wt\%, initial mantle
    temperature $T_\mathrm{init} = 2000$ K, and reference viscosity
    $\mu_{ref} = 1.3 \times 10^{20}$ Pa$\cdot$s are shown: (A)
    variations in HPE budget; (B) variations in mantle reference
    viscosity; (C) variations in initial mantle temperature; and (D)
    variations in mantle C concentration. For all figures, solid lines
    denote an atmospheric loss rate of 30 kg/s, and dashed lines a
    loss rate of 300 kg/s. Dotted line shows 10 bar upper bound on
    LHS~3844b's atmosphere size based on observational constraints.}
  \label{fig:cumulative}
\end{figure*}

The models track the evolution of atmosphere size assuming that
volcanic degassing of CO$_2$, as calculated from the thermal evolution
model, determines the rate at which mass is added to the atmosphere,
and that the rate at which atmosphere is lost is determined by a
constant stripping rate. The erosion of planetary atmospheres is an
area of growing research efforts, particularly in the environment of
low-mass stars where periods of sustained stellar activity can occur
\citep{lammer2008,rodriguezmozos2019}. Based on the stellar activity
discussion in Section~\ref{activity}, we adopt the range of
atmospheric loss rates scaled by the loss rate from Proxima Cen~b of
between 30--300~kg/s \citep{dong2017a}. The atmospheric loss rate in a
particular model is held fixed in time. As atmospheric erosion cannot
occur if no atmosphere is present, we scale the atmospheric loss rate
linearly with atmosphere size when the atmosphere mass is
$<$~44,000~kg (or $< 10^6$ moles of CO$_2$); above this threshold,
atmospheric loss rate is constant.

We also assume a mantle reference viscosity (the viscosity of the
mantle at the present day interior temperature of the Earth) of
$10^{20}-10^{22}$ Pa$\cdot$s, bracketing the range of typical
estimates for the Earth. The radioactive heat budget of the planet has
an important control on thermal evolution and outgassing
\citep{foley2018a}. We consider a range of 50\%--200\% of the Earth's
heat-producing element budget, based on the observational bounds of
radionuclides from stellar abundances
\citep{unterborn2015,botelho2019}. Finally, we also consider a range
of initial mantle potential temperatures of 1700--2000~K. For each of
the three end-member interior structures, 1 million models are run (3
million models in total). The models sample from uniform distributions
of total C budget, atmospheric loss rate, mantle reference viscosity,
heat-producing element budget, and initial mantle
temperature. Logarithmic distributions are used for C budget, loss
rate, and reference viscosity.


\section{Model Results}
\label{results}

Most models show rapid degassing of CO$_2$ within $\sim 1$ Gyrs
(Figure \ref{fig:cumulative}). During this early outgassing stage, the
rate of outgassing far exceeds the atmospheric loss rate, so
atmospheric size rapidly increases. Typically, nearly all of the
planet's supply of CO$_2$ is outgassed during this time, after which
growth of the atmosphere essentially stops; atmospheric stripping then
dominates and the size of the atmosphere shrinks for the rest of the
planet's lifetime. Rapid early outgassing is a result of extensive
early volcanism, which is expected for stagnant lid planets due to
high initial rates of internal heat production and primordial heat. As
LHS~3844b almost certainly lacks surface water, CO$_2$ cannot be
removed from the atmosphere by weathering and the formation of
carbonate rocks. This means there is no mechanism for returning C from
the atmosphere to the mantle, and outgassing irrevocably depletes the
mantle of C. Once the supply of C in the mantle has been depleted,
degassing rates become negligible, even as volcanism continues, and
atmospheric growth stops. In the large majority of our models, mantle
volcanism continues for more than $\sim$1~Gyr, typically lasting
$\sim$3--5~Gyrs.

As most models outgas nearly all of their CO$_2$ to the atmosphere (as
in the examples shown in Figure \ref{fig:cumulative}), the size of
atmosphere formed is largely controlled by the mantle C budget. With
$C_\mathrm{conc} \sim 10^{-4}$ wt\% (2 orders of magnitude smaller
than Earth's C budget in terms of mantle wt\%), the peak atmosphere
size that forms is $\approx 2$ bar; with $C_\mathrm{conc} \sim
10^{-3}$ wt\% (1 order of magnitude smaller than Earth), peak
atmosphere size is $\approx 20$ bar, and with $C_\mathrm{conc} \sim
10^{-2}$ wt\% (approximately equal to Earth's C budget), it is
$\approx 200$ bar. High atmospheric stripping rates of 300 kg/s can
remove enough atmosphere to bring a $\sim$20~bar atmosphere that forms
early in the planet's history down to $< 10$ bar after
$\sim$8~Gyrs. Significantly higher loss rates would be required to
remove the $\sim$200~bar of atmosphere that forms at the upper end of
mantle C budgets we explore. Geophysical factors, such as
heat-producing element budget, mantle viscosity, and initial mantle
temperature, do not have a significant influence on the size of
atmosphere formed or size of atmosphere remaining at LHS~3844b's
current age. Lowering initial mantle temperature or increasing the
reference viscosity can delay the onset of volcanism, leading to
different histories in the first few hundred million years of
evolution, but atmosphere size converges in these models at later
times (Figures \ref{fig:cumulative}B \& C). However, there are
combinations of radionuclide budget, mantle reference viscosity, and
initial temperature that can significantly limit, and even entirely
prevent, mantle degassing as discussed later in this section.

Our estimates of atmosphere size are generally consistent with
\citet{dorn2018b}. \citet{dorn2018b} estimate an atmosphere of
$\approx 70$ bar for a stagnant lid planet of similar mass to
LHS~3844b. Their models assume 1000 ppm as the concentration of CO$_2$
in the mantle, which corresponds to $\sim$$10^{-3}$~wt\%, the middle
of the range of CO$_2$ concentrations we consider; with this
concentration, we estimate a comparable CO$_2$ atmosphere size of
$\approx 20$ bar. \cite{dorn2018b} also keep the concentration of
CO$_2$ in the mantle fixed over time, rather than treating it as a
CO$_2$ reservoir that shrinks due to outgassing, though they do track
mantle depletion and assume depleted regions have their volatile
abundances lowered by melting. This difference may explain the larger
CO$_2$ atmosphere they estimate ($\approx 70$ bar versus $\approx 20$
bar).

\begin{figure*}
  \begin{center}
    \includegraphics[angle=270,width=17cm]{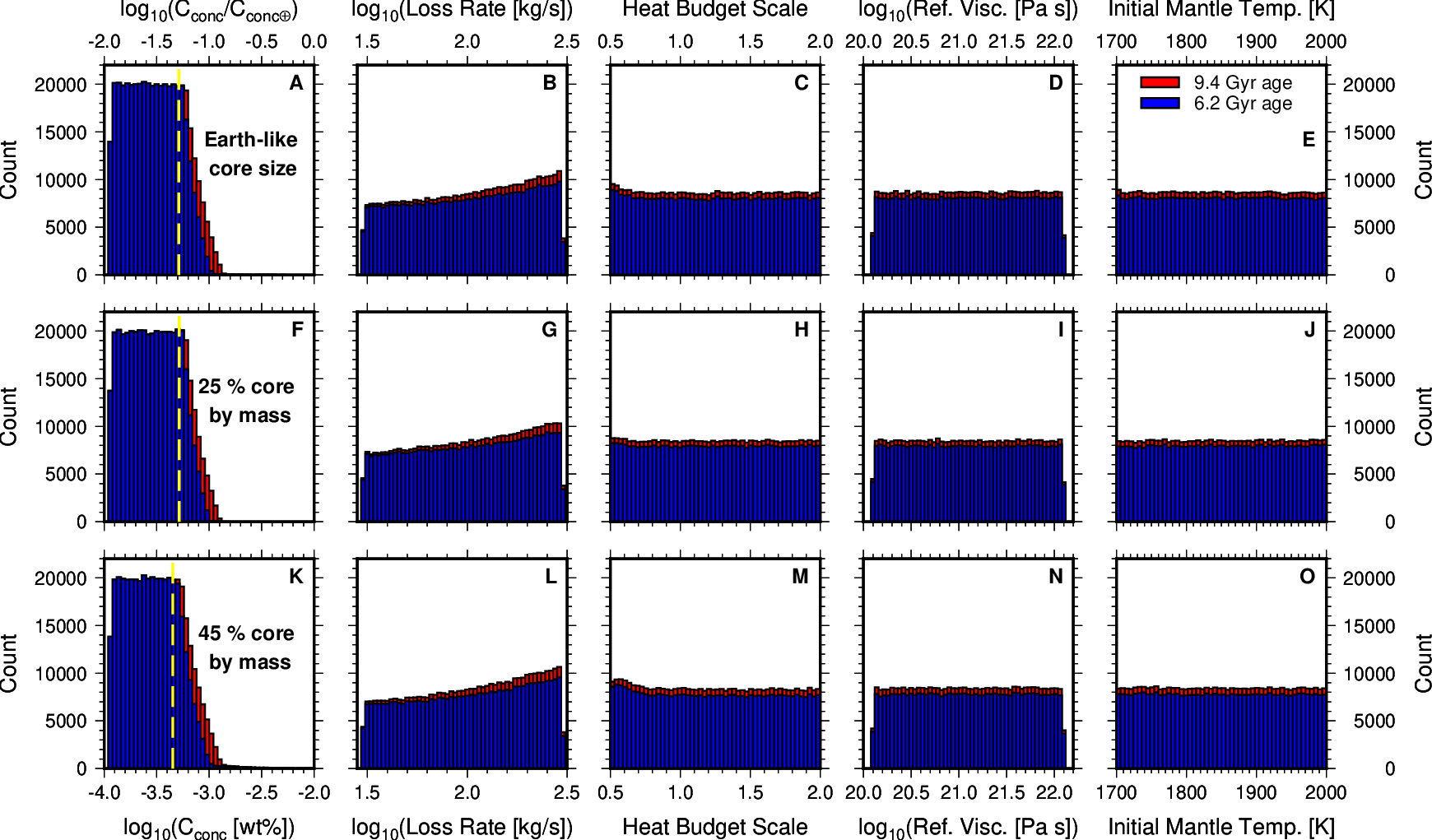}
  \end{center}
  \caption{Histograms of all model results that successfully produce
    an atmosphere of $<$ 10 bar after 6.2 Gyrs (blue) and 9.4 Gyrs
    (red). Top row shows models with an Earth-like core size, middle
    row models with a small core, and bottom row models with a large
    core. Shown are the number of successful models as a function of
    the mantle C budget (A, F, \& K), the atmospheric loss rate (B, G,
    \& L), the heat-producing element budget relative to Earth's (C,
    H, \& M), the mantle reference viscosity (D, I, \& N), and the
    initial mantle temperature (E, J, \& O). Dashed lines show the
    limit where the mantle C budget is low enough that even complete
    outgassing of all mantle C will result in a $< 10$ bar atmosphere
    (A, F, \& K). Plots of mantle C budget are presented in terms of
    mantle C concentration, $C_\mathrm{conc}$, in wt\% (lower axis)
    and mantle C concentration normalized to Earth,
    $C_\mathrm{conc}/{C_\mathrm{conc}}_\oplus$ (upper axis).}
  \label{fig:all_hists}
\end{figure*}

The examples shown in Figure~\ref{fig:cumulative} demonstrate that
atmospheric loss rate and mantle C budget have a significant control
over the ending atmosphere size. To elucidate the full combination of
model parameters that result in atmospheres fitting the observational
constraints for LHS~3844b, we show histograms of all successful models
in Figure \ref{fig:all_hists}. There is a clear trend of low C budgets
and high loss rates being able to explain the thin atmosphere. For all
three assumed interior structures we modeled, $C_\mathrm{conc} <
\approx 5 \times 10^{-4}$ wt\% results in a large number of successful
models; the number of successful models then decreases rapidly from
$C_\mathrm{conc} \approx 5 \times 10^{-4} - 10^{-3}$ wt\%, and very
few successful models result from $C_\mathrm{conc} > \sim 10^{-3}$
wt\%. Thus models with C budgets of approximately an order of
magnitude lower than Earth's or larger almost always produce
atmospheres larger than the observational constraint.

Successful models are found for all modeled atmospheric loss rates,
but higher loss rates clearly lead to a higher proportion of
successful models. Successful models are found in near-uniform
distributions of heat-producing element budget, mantle reference
viscosity, and initial mantle potential temperature. There is a small
increase in the proportion of successful models with low
heat-producing element budgets. This indicates that producing an
atmosphere $< 10$ bar is not sensitive to initial mantle temperature
or reference viscosity, and largely insensitive to radionuclide
budget, save for slightly favoring low radionuclide budgets. An
important feature of the results is that the assumed interior
structure does not have a significant impact; the Earth-like, small
core, and large core cases all show very similar distributions of
successful models among the considered parameters. Whether the low or
high end of the estimated age range for LHS~3844b is used also does
not have a significant effect, though intuitively, more models are
successful when the upper bound on age is used.

The influence of mantle C budget and loss rate can be easily
explained. First, there is a limit below which there is simply not
enough C in the mantle to produce an atmosphere $> 10$ bar. Below this
limit, all models will be successful, as it will not be possible to
form an atmosphere above the current observational constraints. As the
atmospheric pressure is given by
\begin{equation}
    P_\mathrm{atm} = \frac{C_\mathrm{atm} m_{CO_2} g}{A_s}
\end{equation}
where $C_\mathrm{atm}$ is the number of moles of CO$_2$ in the
atmosphere, $m_{CO_2}$ is the molar mass of CO$_2$, and $A_s$ is the
surface area of the planet, then for
\begin{equation}
    C_\mathrm{tot} < \frac{(10 \mathrm{bar}) A_s }{m_{CO_2}g}
\end{equation}
all models will be successful; this limit is found to be $1.25 \times
10^{21}$ mol ($C_\mathrm{conc} = 5.14 \times 10^{-4}$ wt\%) for an
Earth-like core, $1.37 \times 10^{21}$ mol ($C_\mathrm{conc} = 5.18
\times 10^{-4}$ wt\%) for the small core case, and $1.04 \times
10^{21}$ mol ($C_\mathrm{conc} = 4.50 \times 10^{-4}$ wt\%) for the
large core case. These limits are shown in Figure \ref{fig:all_hists}
as vertical dashed lines.

\begin{figure*}
  \begin{center}
    \includegraphics[angle=270,width=16cm]{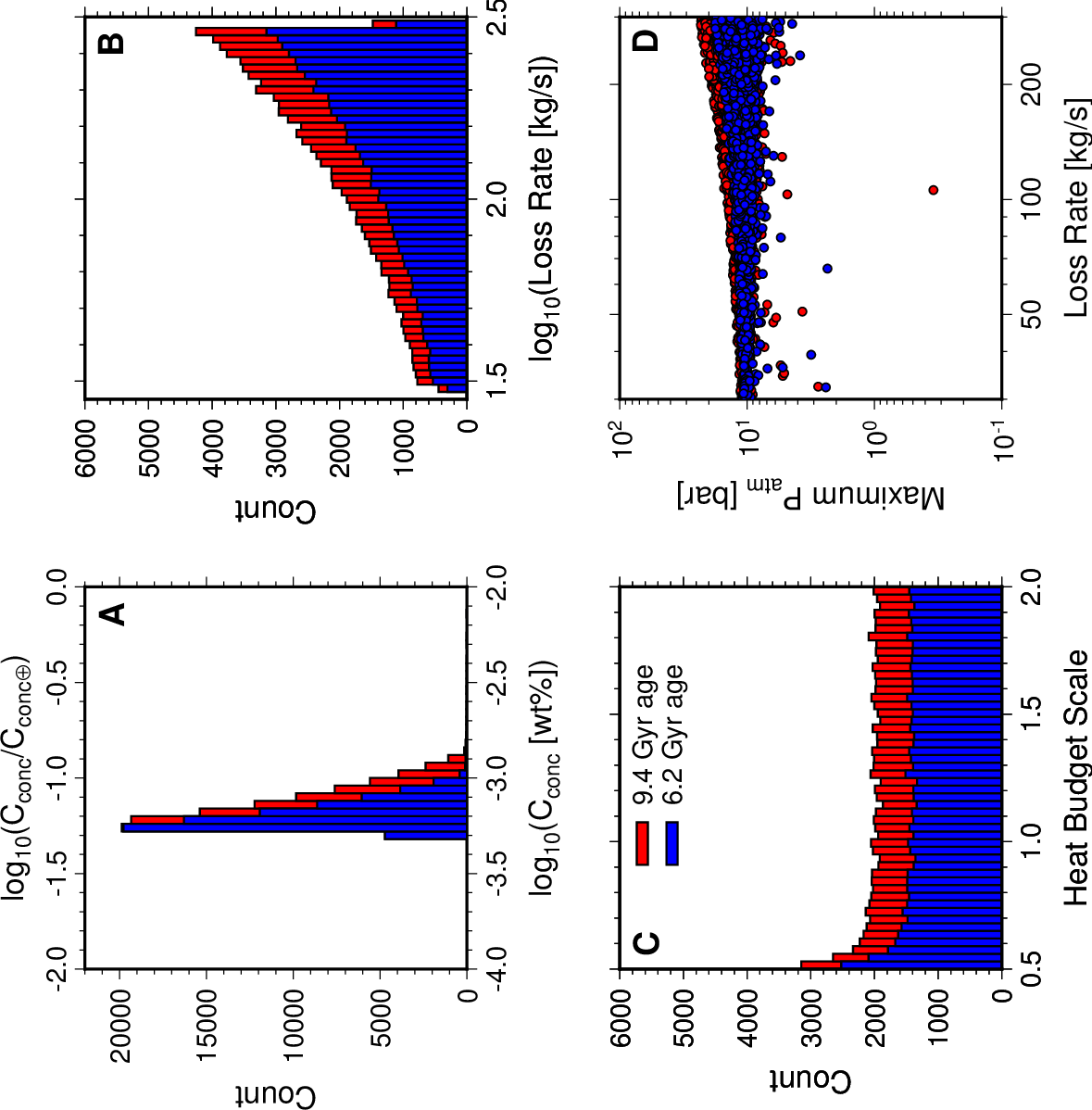}
  \end{center}
  \caption{Successful models with an Earth-like core size and a mantle
    C budget $>5.14 \times 10^{-4}$ wt\%; that is, enough CO$_2$ to
    form an atmosphere larger than 10 bar. Histograms of these
    successful models as a function of mantle C budget (A),
    atmospheric loss rate (B), and heat-producing element budget
    relative to the Earth (C) are shown; blue bars are for an age of
    6.2 Gyrs, red for an age of 9.4 Gyrs. Also shown is the maximum
    atmospheric pressure reached during the model runs, as a function
    of the atmospheric loss rate (D). Only every 50th model run is
    shown in this panel, to make the figure more readable.}
  \label{fig:ctot_lim}
\end{figure*}

For larger $C_\mathrm{tot}$ (or $C_\mathrm{conc}$), an atmosphere $>
10$ bar can form if significant outgassing occurs. Thus, atmospheric
stripping is required to bring the atmosphere back to within
observational constraints. Histograms of all successful models with
$C_\mathrm{conc} > 5.14 \times 10^{-4}$ wt\% for the Earth-like core
size case are shown in Figure \ref{fig:ctot_lim}; similar
distributions are seen for the small core and large core models (not
shown). There is a clear tradeoff between $C_\mathrm{conc}$ and loss
rate. Higher loss rates allow larger $C_\mathrm{conc}$ values to still
satisfy the atmosphere size constraint, while lower $C_\mathrm{conc}$
values allow even low atmospheric loss rates to still result in
successful models. Specifically, for a given loss rate, there is a
maximum atmospheric size that can be formed early in the planet's
history and still be stripped to $< 10$ bar after 6.2--9.4~Gyrs. This
limit on maximum atmospheric size is illustrated in Figure
\ref{fig:ctot_lim}D, which shows the peak atmospheric size reached
during each successful model run as a function of loss rate. For a
loss rate of 300 kg/s, atmospheres can only reach $\approx$20--25~bar
and still be stripped to $< 10$ bar within LHS~3844b's lifetime. Our
assumed upper limit on atmospheric loss rate is simply not high enough
to remove atmospheres larger than 20--25 bar.

In fact, we can still make a simple estimate of this maximum
atmospheric pressure ($P_\mathrm{atm}^\mathrm{max}$) that can be
reached and still satisfy the constraint on present-day atmosphere
size. Assuming that the atmosphere forms very early in the planet's
history, as seen in our outgassing models, the total mass of
atmosphere that can be stripped is given by the product of the loss
rate ($E$) and the planet age ($\tau$); thus
\begin{equation}
    P_\mathrm{atm}^\mathrm{max} = 10 \mathrm{bar} + \frac{E \tau g}{A_s} .
\end{equation}
for $E=300$~kg~s$^{-1}$, $\tau = 9.4$ Gyrs, and $g=15.7$~m~s$^{-2}$, as
for an Earth-like core size, $P_\mathrm{atm}^\mathrm{max} \approx 25$
bar, nearly identical to the upper limit seen in Figure
\ref{fig:ctot_lim}D. Further assuming that all of the planet's mantle
CO$_2$ budget is outgassed, the upper bound on mantle C budget that
can explain the present-day state of LHS~3844b's atmosphere, as a
function of atmospheric loss rate, can be calculated as
\begin{equation}
  C^{max}_{tot} = \frac{10 \mathrm{bar} \times A_s}{m_{CO_2}g} +
  \frac{E \tau}{m_{CO_2}}.
  \label{eq:c_tot_max}
\end{equation}
With the same numbers as listed above, $C_\mathrm{tot}^\mathrm{max}
\approx 3 \times 10^{21}$ mol, which corresponds to a mantle C
concentration of about 10\% Earth's. For LHS~3844b to have an
Earth-like mantle C budget, the atmospheric stripping rate would have
to be a factor of 10 larger than our estimated upper bound.

\begin{figure*}
  \begin{center}
    \includegraphics[angle=270,width=17cm]{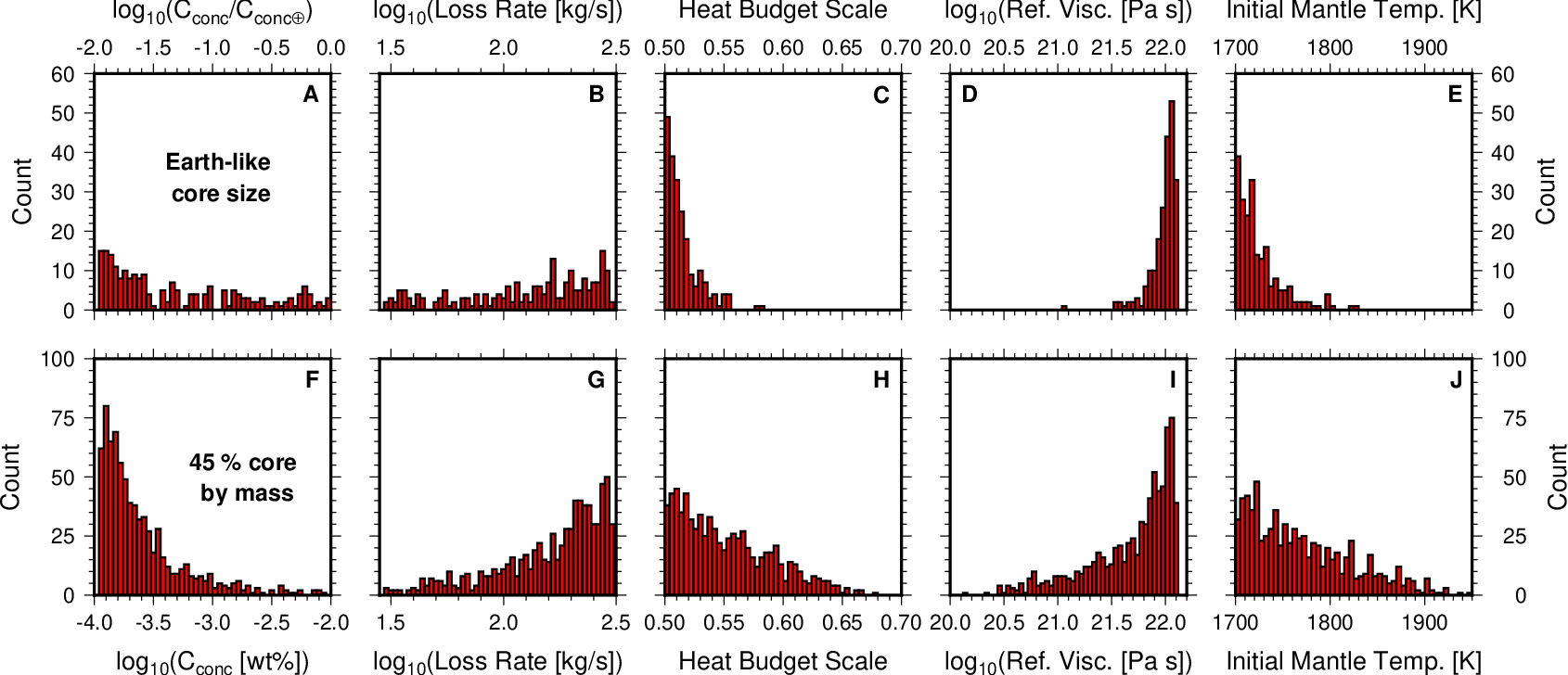}
  \end{center}
  \caption{Histograms as in Figure \ref{fig:all_hists}, except only
    models that result in a maximum atmospheric pressure during the
    model run of $< 10^{-5}$ Pa are shown. These are models that
    experience virtually no outgassing, due to a lack of
    volcanism. Only the Earth-like core and large core model suites
    are shown, as with a small core nearly all models result in
    significant volcanism and outgassing. }
  \label{fig:no_outgas}
\end{figure*}

Finally, Figure~\ref{fig:ctot_lim}C shows that more models are
successful when the heat-producing element budget is low. This reveals
one additional group of models that are able to produce atmospheres of
$< 10$ bar; models with very limited (or even completely absent)
volcanic outgassing. To illustrate the factors that allow planets to
experience limited outgassing, we plot all models that produce maximum
atmosphere pressures $< 10^{-5}$ Pa in
Figure~\ref{fig:no_outgas}. Here, we show only the Earth-like core and
large core models, as very few of the small core models resulted in
such limited outgassing. The cutoff of $10^{-5}$ Pa was used because
there was a clear grouping of models with maximum atmospheric sizes
below this threshold when looking at the entire suite of models
together. Models that experience essentially no outgassing are
characterized by low heat-producing element budgets, low initial
mantle potential temperatures, and high mantle reference
viscosities. All of these factors act to suppress mantle melting. Low
initial temperatures and radionuclide budgets limit mantle heating
and can keep the mantle cool enough to not melt and produce
volcanism. A high reference viscosity increases the thickness of the
stagnant lid, which also suppresses volcanism. A thicker stagnant lid
requires a higher mantle temperature for melting to occur, because it
forces melting to occur at higher pressures. In the large core models,
there is also a trend toward low C budgets and high loss rates, as
these factors can keep the atmospheric size low even if some limited
volcanism occurs. However, it should be noted that volcanism can be
shutoff by purely geophysical factors, mantle heat budget and
viscosity, in which case no atmosphere will form even if the C budget
is very large or loss rate is small.


\subsection{Carbon in the Core}

The thermal evolution models indicate that a mantle depleted in carbon
relative to the Earth is the most likely explanation for the
present-day thin atmosphere of LHS~3844b. However, whether a
C-depleted mantle means the overall planet composition is depleted in
carbon relative to the Earth depends on the partitioning of carbon
between the mantle and core. For a given bulk planet composition of
carbon, the core composition and thus initial mantle concentration are
set as the core segregates during the planet's magma ocean stage
\citep{dasgupta2013,fischer2020}. Geochemical estimates, based on
meteoritic abundances, show the bulk Earth (crust + mantle + core)
contains 0.07 wt\% carbon total \citep{mcdonough2003}. Of this bulk
carbon, the Earth's core contains $\sim$90\% of the total budget,
yielding a core C concentration of 0.2 wt\%. This leaves only 0.01
wt\% C in the mantle \citep{mcdonough2003}. This ratio of the mass of
C split between the mantle and core is known as the partition
coefficient ($D_C = f_\mathrm{C,met}/f_\mathrm{C,mantle}$). These
geochemical models predict $D_C=9$ for the Earth. However, $D_C$ is a
complex function of the pressure, temperature, and oxygen fugacity
($fO_2$) of the magma ocean upon equilibration, the fraction of sulfur
and oxygen in the iron melt, and the degree of silicate melt
polymerization \citep{fischer2020}. We define the affinity for an
element to partition into metallic iron as siderophile. Generally, a
larger fraction of C enters the mantle, that is, C becomes less
siderophile, as pressure, degree of melt polymerization, and the
fraction of S in the melt increase. Conversely, more C enters the
core, that is C becomes more siderophile, as the fraction of O in the
melt and $fO_2$ increase. Core segregation can happen in two ways; as
a single event where material equilibrates at pressures near the
core-mantle boundary and in a multistage fashion where accreting
material slowly adds volatiles to both the core and mantle, leading to
much lower equilibration pressures.

We model the competing effects of pressure, composition, and oxygen
fugacity on C partitioning into the core assuming single-state core
formation. We use the formulation of \citet{fischer2020} for
determining $D_c$, and thus estimate the fraction of C that is
retained in the mantle for a given bulk C budget of LHS 3844b. We
assume a degree of polymerization comparable to peridotite (2.6), and
in order to maximize the fraction of C in the core, set the fraction
of S in the melt to zero. We then construct a model of $D_C$
containing four variables: pressure ($P$), temperature ($T$), the
fraction of O in the metal ($X_{O}^\mathrm{metal}$), and $fO_2$ of the
material when the metal and silicate equilibrate. The value of
$X_{O}^\mathrm{metal}$ is also a function of $P$, $T$ and
$fO_2$. Using the exchange coefficient ($K_D$) formulation as a
function of $P$ and $T$ of \citet{fischer2015} for O, we can estimate
the fraction of O in the melt using the relation: $K_D =
(X_{Fe}^\mathrm{metal} X_{O}^\mathrm{metal}) /
(X_{FeO}^\mathrm{silicate})$. The equilibrium ratio
$X_{Fe}^\mathrm{metal}/X_{FeO}^\mathrm{silicate}$ is determined using
the definition of the iron-w\"{u}stite fugacity buffer that assumes
the activity of Fe and FeO in the system are approximated by their
mole fractions:
\begin{equation}
  \Delta \rm{IW} = 2\log_{10} \left(
  \frac{X_{FeO}^\mathrm{silicate}}{X_{Fe}^\mathrm{metal}} \right).
\end{equation} 
We adopt the liquidus of \citet{stixrude2011} to calculate the minimum
temperature of the magma ocean as a function of pressure.

\begin{figure}
  \includegraphics[width=8cm]{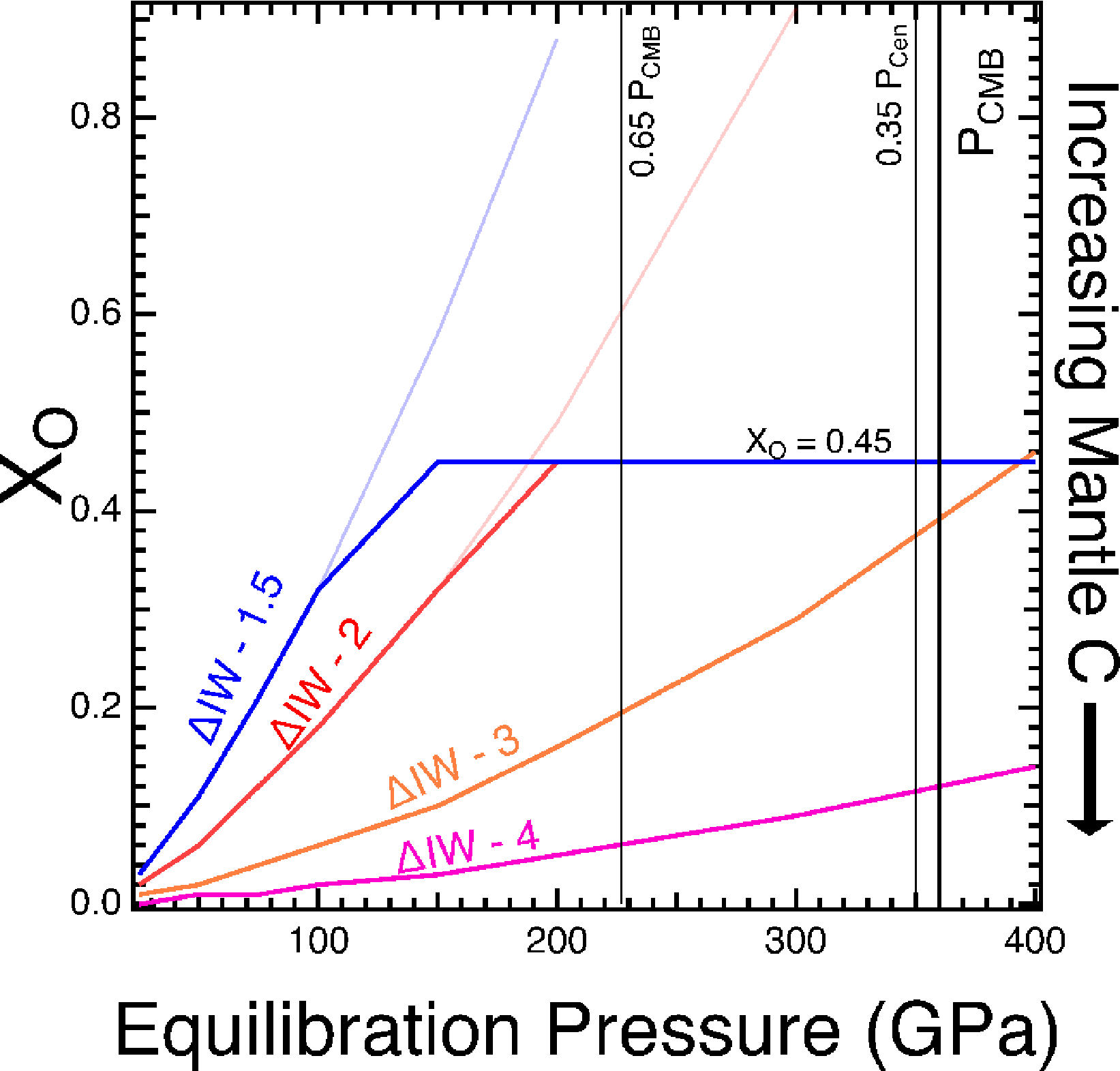} \\
  \includegraphics[width=8cm]{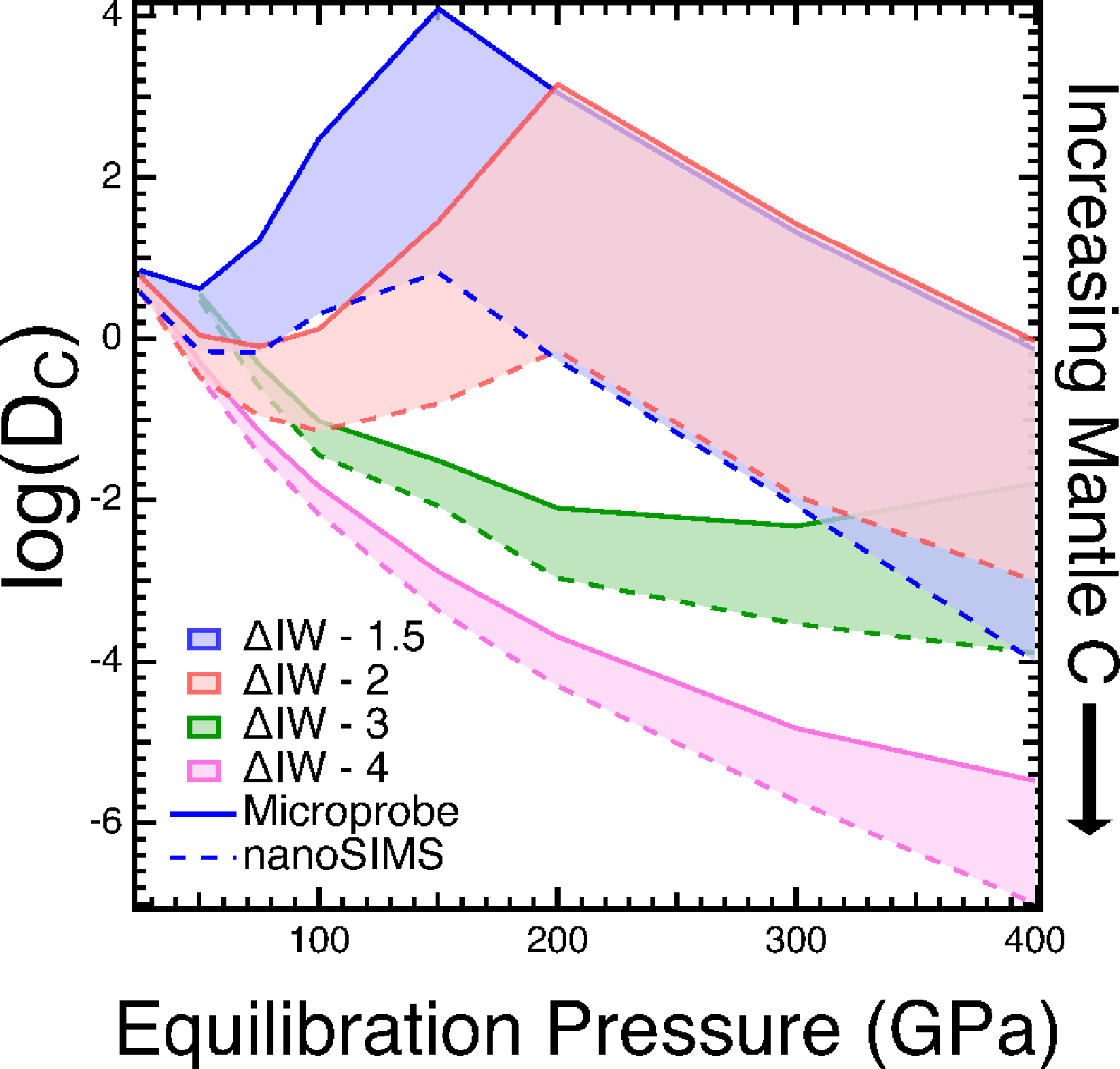}
  \caption{Top: Fraction of oxygen present in the metallic melt as a
    function equilibration pressure and of bulk $fO_2 = \Delta
    \rm{IW}=$ -1.5 (blue), -2 (red), -3 (orange) and -4 (pink). We
    assume a maximum value of $X_O^{metal}$ of 0.45 however
    determinations of $X_O^\mathrm{metal}$ without this assumption are
    shown as thin lines. Bottom: Log of partition coefficient of C
    ($\log(D_C$)) calculated from nanoSIMS (dashed) and microprobe
    data (solid) as a function of bulk $fO_2$ of the planet during
    core equilibrium and segregation. We adopt the average of these
    two values for our models; however, this model assumes only Fe, C,
    and O are present in the system, and it lacks Si. As Si enters
    into the core, $X_{O}^\mathrm{metal}$ will decrease, and therefore
    we consider these upper limits at all pressures.}
    \label{fig:fracO}
\end{figure}

The value of $X_{O}^\mathrm{metal}$ increases with increasing
equilibration pressure (Figure \ref{fig:fracO}, top). We note that the
highest pressure for measurements in \citet{fischer2015} is
$\sim$100~GPa at $\Delta \rm{IW} = -1.1$, which estimates
$X_{O}^\mathrm{metal}$ of 0.27. For $fO_2 > \Delta \rm{IW}-2)$ and
high pressure, $X_{O}^\mathrm{metal}$ rapidly increases, likely due to
our assumption that $X_{FeO}^\mathrm{silicate} /
X_{Fe}^\mathrm{metal}$ approximates their activities at high pressure
and temperature. Additionally, our simple model only assumes
equilibration in the Fe-C-O system. Importantly missing from this
model is the partitioning of Si between the mantle and core during
formation. The relative fractions of Si and O entering the core are
inversely correlated \citep{fischer2015}, with relatively more Si
entering the core under reducing conditions. As such, we likely
overestimate $X_{O}^\mathrm{metal}$ and $D_C$ at high pressures and
temperatures, particularly under oxidizing conditions where our model
predicts $>80$\% of the mass of metal will be O. We therefore set an
arbitrary maximum value of $X_{O}^\mathrm{metal}$ = 0.45 and will
explore the consequences of relaxing this assumption below.

From $X_{O}^\mathrm{metal}$, we can derive the partition coefficient
of C ,$D_C$, as a function of pressure and the $fO_2$ of the magma
ocean (Figure \ref{fig:fracO}, bottom). \citet{fischer2020} provides
two determinations of $D_C$ produced using either fits to nanoSIMS- or
electron microprobe-derived datasets, with the $D_C$ values from the
microprobe fit being higher than the nanoSIMS fit(Figure
\ref{fig:fracO}, bottom). For oxidizing magma ocean conditions ($fO_2
> \Delta \rm{IW}-2$), $D_C$ increases until between 150 and 200~GPa,
where it begins to drop as the effect of increasing pressure lowering
$D_C$ begins to dominate due to our arbitrary maximum value of
$X_{O}^\mathrm{metal}$ limiting the amount of O (and thus C) available
to partition into the core. Under reducing conditions,
$X_{O}^\mathrm{metal}$ remains low and $D_C$ decreases over all
pressures.

\begin{figure*}
  \begin{center}
    \begin{tabular}{ccc}
      \includegraphics[width=5.5cm]{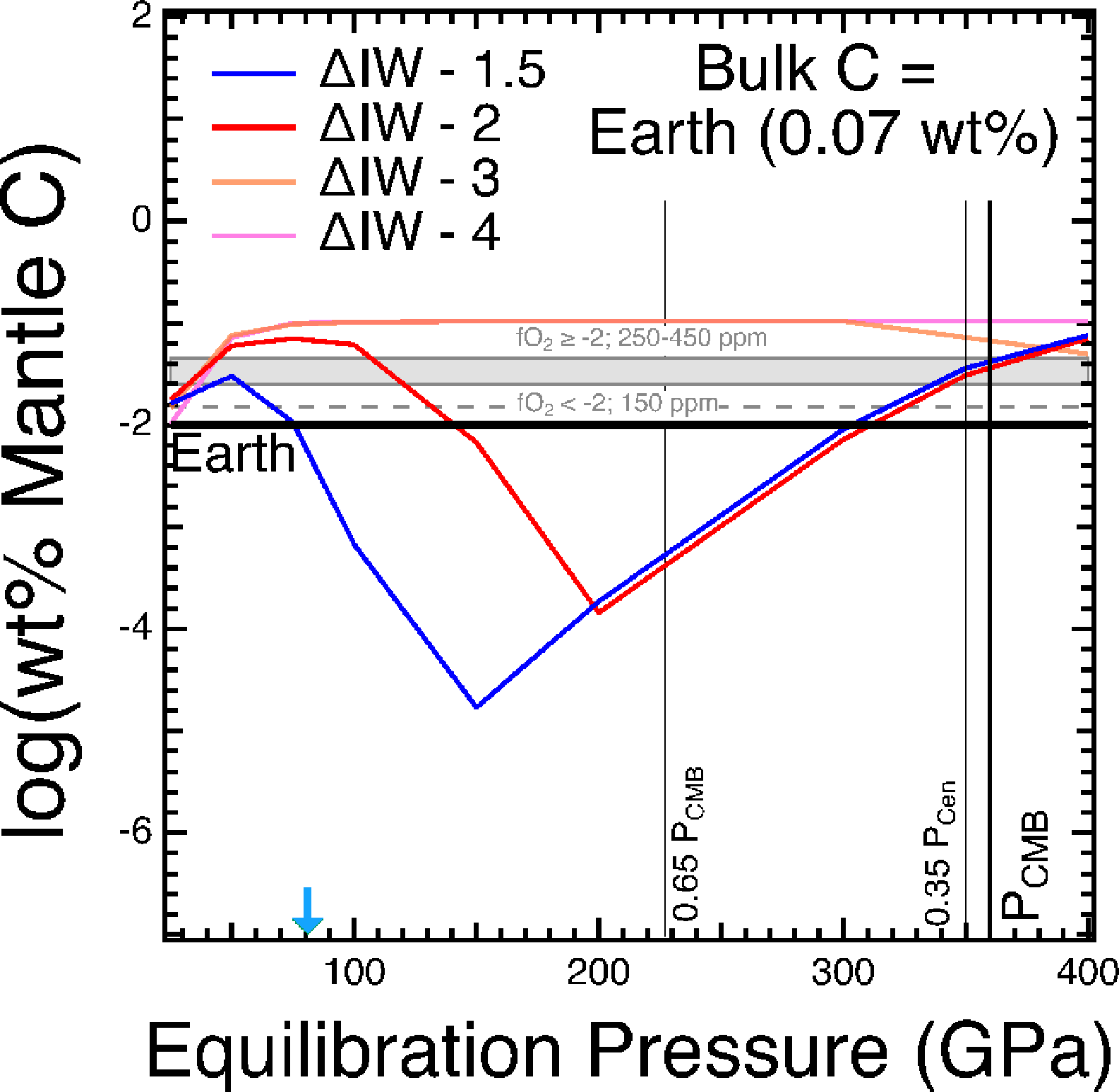} &
      \includegraphics[width=5.5cm]{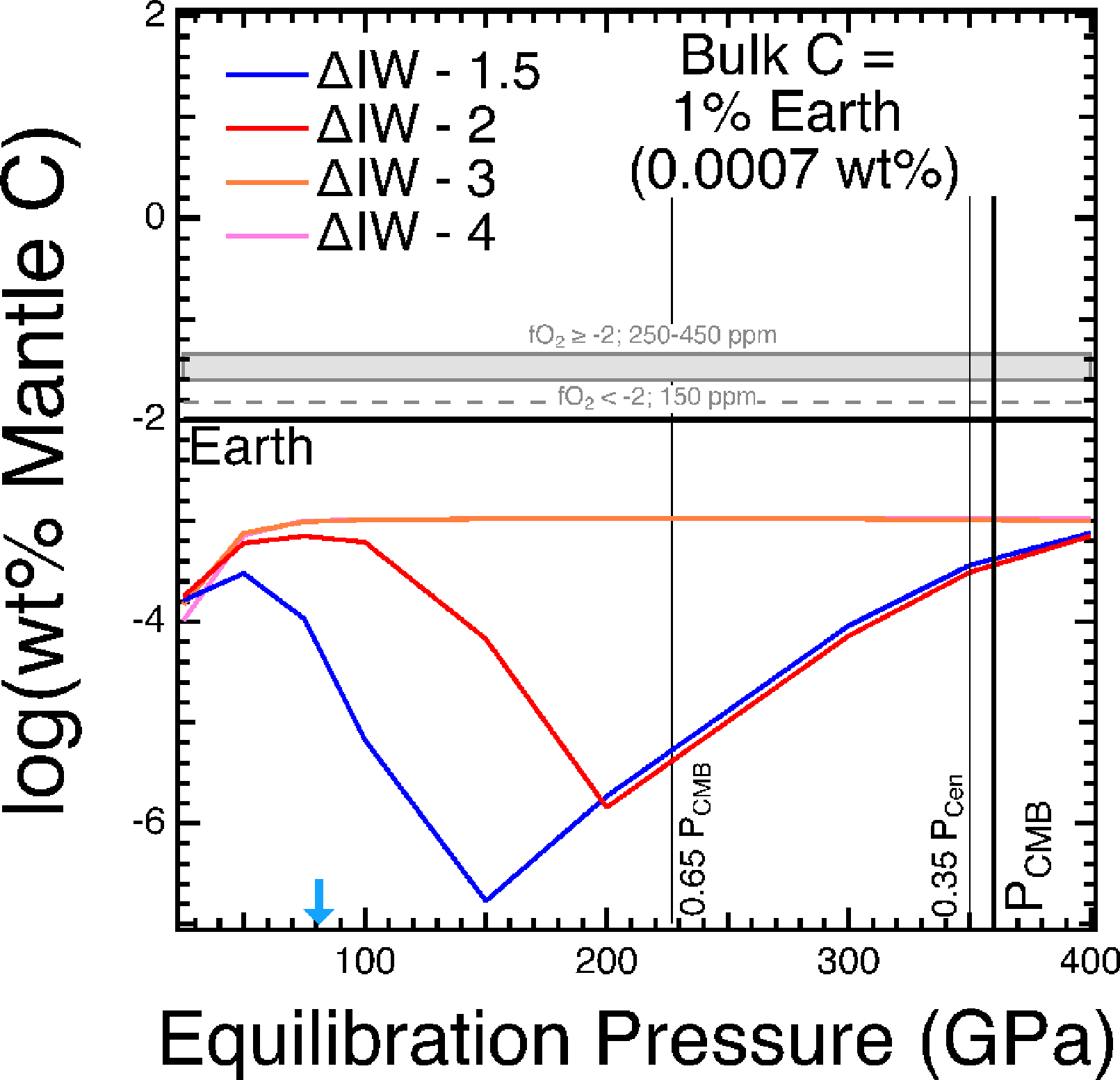} &
      \includegraphics[width=5.5cm]{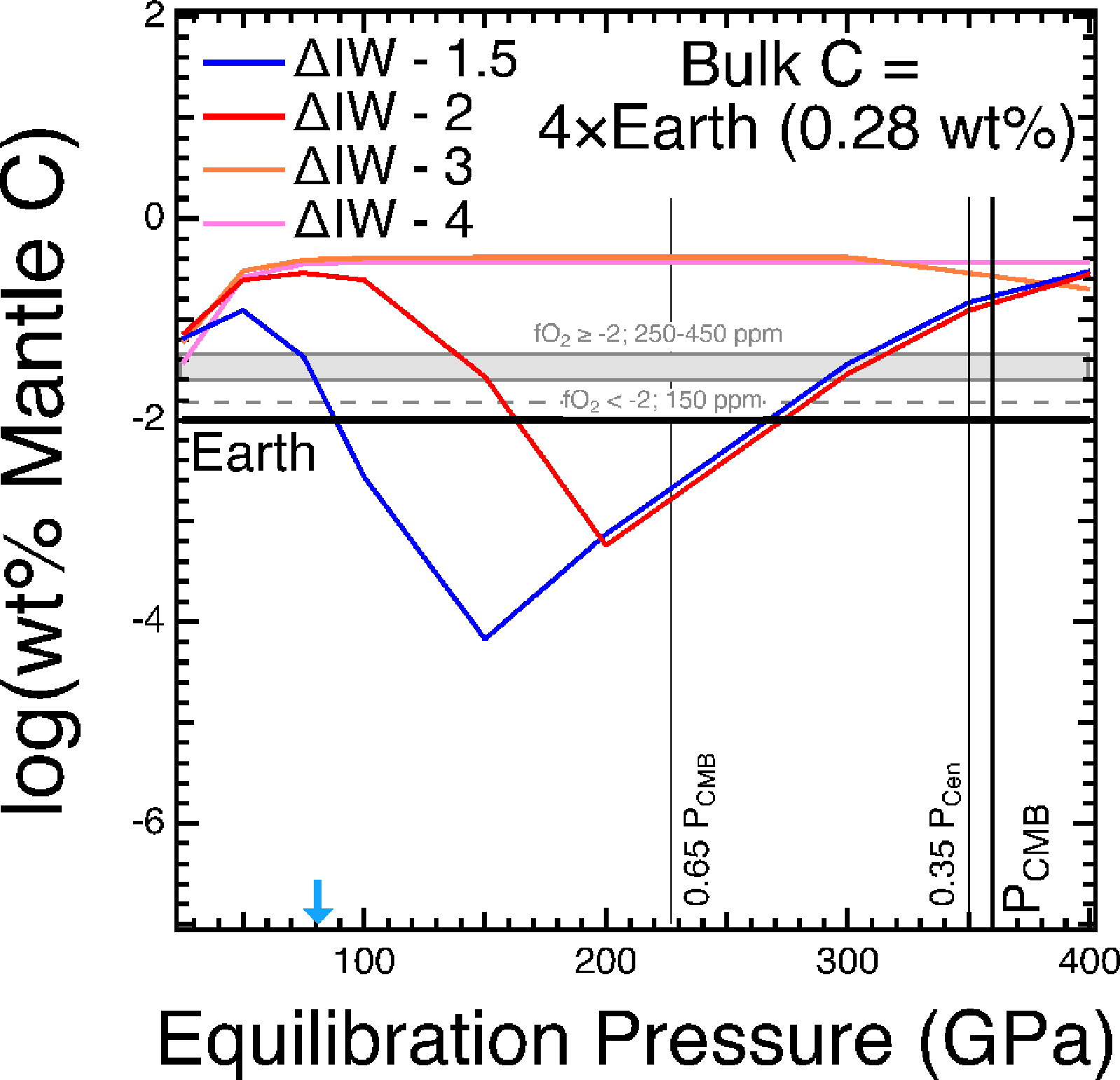}
    \end{tabular}
  \end{center}
  \caption{Left: Resulting mantle C concentrations calculated assuming
    LHS~3844b has an Earth-like core (CMF = 0.33), Earth concentration
    of C (0.07 wt\%) and equilibrated at fO$_2$ = $\Delta$Iw - 1.5
    (blue), $\Delta$Iw - 2 (red), $\Delta$IW - 4 (pink) and $\Delta$IW
    - 6 (orange). Solid curves represent increasing
    $X_{O}^\mathrm{metal}$ as pressure increases, adopting the $K_D$
    values of \citet{fischer2015} and dashed lines assume a maximum
    value of $X_{O}^\mathrm{metal}$ of 0.4. Center: Same as left
    assuming an planet with 1\% of Earth's C concentration (0.0007
    wt\%). Right: Same as left assuming a planet with $4\times$
    Earth's C concentration (0.28 wt\%). For all figures, horizontal
    lines mark the Earth's mantle C-content \citep[solid black;
    ][]{mcdonough2003} and the solubility of C in the mantle at 3~GPa
    from \citet{grewal2020} for reducing conditions of 150 ppm ($fO_2
    < \Delta \rm{IW}-2$; gray dashed) and for oxidizing conditions of
    between 250--450~ppm ($fO_2 \geq \Delta \rm{IW}-2$; gray
    box). Vertical lines mark the max pressure for equilibration at
    the CMB ($P_{CMB} = 360$~GPa) as well as the single-core formation
    equilibration pressure that is 35\% of the central core pressure,
    both taken from \citet{unterborn2019}. The Earth's core
    equilibration pressure under multi-stage core formation
    \citep{fischer2017} is shown as a teal arrow, for reference.}
  \label{fig:core}
\end{figure*}

Under single stage core formation, equilibration occurs at roughly
35\% of the central pressure of the planet \citep{schaefer2017a}. For
LHS-3844b, we estimate this occurs at $\sim$350~GPa using the pressure
estimates of a 1.3 $R_\oplus$ planet from
\citet{unterborn2019}. Adopting the average $D_C$ of the microprobe
and nanoSIMS fits, we estimate that if LHS~3844b formed with an
Earth-like concentration of C (0.07 wt\%) and an Earth-like core mass
fraction (0.33), that under oxidizing conditions ($\Delta \rm{IW} \geq
-2$), little C is sequestered into the core producing a mantle with
0.03 wt\% C, a factor of 3 greater than the Earth's (Figure
\ref{fig:core}, left). Interestingly, if core equilibration happened
between 80--280~GPa under these oxidizing conditions, significant
amounts of C would partition into the core, leaving behind a mantle
significantly depleted in C, akin to what has been found for
Earth-size planets \citep{li2015,li2016}. Under reducing conditions
($\Delta \rm{IW} < -2$), $D_C$ is low and practically no carbon is
partitioned into the core, producing a mantle with an order of
magnitude greater C concentration than the Earth. Exploring the
effects of bulk C budget, we find that if LHS~3844b formed with 1\% of
the Earth's C budget (0.0007 wt\%), regardless of the oxidation state
of the magma ocean, the mantle will have a concentration below that of
the Earth, whereas for a planet with four times the C budget of Earth
(0.28 wt\%), the mantle will have an even greater C concentration than
our Earth concentration model does (Figure~\ref{fig:core}, center,
left). These relatively simple models show that, under single-stage
core formation, the mantle's C concentration is primarily a function
of the total amount of C present in the planet. In order to produce a
mantle with a C budget low enough to result in a $\leq 10$ bar
atmosphere for LHS~3844b (0.002 wt\%, Figure~\ref{fig:all_hists}), the
planet overall would need to be depleted in C relative to the Earth.

Under single-stage core formation little C enters LHS~3844b's core,
regardless of magma ocean oxidation state. For those models where bulk
is at or higher than Earth's, this lack of C partitioning into the
core will produce a magma ocean at carbon saturation. When a magma
ocean is at saturation, some fraction of the C will dissolve into the
silicate melt, with the rest likely becoming outgassed. Under reducing
conditions ($fO_2 < \rm{IW}-2$) the solubility of C in a silicate melt
at the low pressures ($P \sim 3$~GPa) where degassing occurs is
$\sim150$~ppm \citep[][, Figure~\ref{fig:core}, gray dashed
  line]{grewal2020}. Our reduced Earth-composition model of LHS~3844b
predicts a mantle concentration of $\sim1000$ ppm
(Figure~\ref{fig:core}, left), whereas the low C-content model will
retain all of its C in the mantle, likely as reduced C in the form of
diamond or graphite \citep{unterborn2014}. Under oxidizing conditions
($fO_2 > \Delta \rm{IW} -2$), C solubility in silicate melt increases
to $\sim$250~ppm at $fO_2 = \Delta \rm{IW}-2$ and 450~ppm for $fO_2 =
\Delta \rm{IW} -1.5$ at these pressures (Figure~\ref{fig:core}, gray
box). In this case, only the model with four times the bulk C budget
of the Earth will produce a mantle above this solubility range. In
both the oxidizing and reducing cases, for those mantle contents above
their respective solubilities, mantle degassing of C-bearing gas will
occur until mantle C contents become comparable to the
solubility. These degassed proto-atmospheres will be comprised of
CO$_2$ and CH$_4$ for magma oceans with oxidizing and reducing $fO_2$,
respectively \citep{kasting1993e,gaillard2009,grewal2020}.  Even after
the considerable degassing, the final mantle composition will still
remain above the Earth's (Figure~\ref{fig:core}, thick black line).

Under single-stage core formation, the high pressure of equilibration
prevents C from partitioning into the core. This means a planet's bulk
C budget coupled with the solubility of C in the silicate melt
effectively sets the initial mantle C budget and whether or not a
proto-atmosphere will be created. Notably, this proto-atmosphere must
also be eroded to explain LHS~3844b's current lack of
atmosphere. Under reduced core formation scenarios, $D_C$ decreases
with increasing pressure, meaning C-rich mantles comprised of
silicates with graphite/diamonds and a CH$_4$ atmosphere are the
likely initial state of rocky super-Earths after magma ocean
crystallization \citep{grewal2020}. Under oxidizing conditions,
however, only if $X_O^\mathrm{metal}$ is greater than our arbitrary
value of 0.45 will $D_C$ increase at high pressure, allowing for more
C to partition into the core. No experiments in the Si-O-Fe-C system
have been performed above $\sim$60~GPa, and thus we are extrapolating
well beyond the range of experimental data. Regardless, to increase
$X_{O}^\mathrm{metal}$ values above 200 GPa, $K_D$ must also
increase. Otherwise, under single-stage core formation, C is simply not
siderophile enough at oxidizing fugacities to partition significant
amounts of C into the core.  We note, however that the primary oxidant
in this case would be water. Much like CO$_2$ or CH$_4$, the amounts
of water needed to produce these oxidizing $fO_2$s will also produce
significant water vapor and surface oceans \citep{eilkinstanton2011b},
which must also be eroded in addition to the CO$_2$ to explain
LHS~3844b's current lack of atmosphere.

Multi-stage core formation, where accreting material equilibrates with
the magma ocean, is another viable core formation scenario
\citep{rubie2011,fischer2017,fischer2020}. In multi-stage core
formation, the planet's core and mantle grow via a series of
equilibration processes, as infalling material equilibrates at low
pressures initially when the planet is small, and equilibration
pressure increases as the planet grows. This allows for core formation
to occur at much lower pressures than single-stage core formation,
with equilibration pressures reaching a maximum of $\sim$65\% of the
CMB pressure that would occur if it formed as a single event
\citep{rubie2011,fischer2017}. For the Earth, this pressure is
$\sim$80 GPa compared to its CMB pressure of $\sim$125 GPa. For
LHS~3844b, 65\% $P_{CMB}\sim$230~GPa (Figure~\ref{fig:core}). Under
oxidizing conditions, as material is added and the pressure of
equilibration increases, O and C become more siderophile. As such, the
C of early infalling material will equilibrate at low pressures and
its C will stay in the mantle. Later infalling material, however, will
equilibrate at higher pressures and the majority of the accreting
material's complement of C will partition into the core. This scenario
is only likely, though, if LHS~3844b formed entirely out of oxidized
material. The Earth, however, likely formed from a mixture of
oxidizing and reduced material, which would explain its trace element
abundances \citep{wade2005}. If LHS~3844b also formed from a mixture
of material of varying $fO_2$, the mantle may still be left C-rich due
to the low $D_C$ of the reducing material, despite some fraction of
the planet's C being partitioned into the core as a consequence of
accreting oxidizing material as well. Whether the balance of the low
pressure of equilibration in multi-stage formation and the planet
forming from a mixture of reduced and oxidizing material are able to
produce C-poor mantles for planet's with Earth-like or greater C
budgets is beyond the scope of this paper. In the absence of new
experimental data for the partitioning of O, Si and C at $P >
150$~GPa, in both single-stage core formation and multi-stage core
formation, our models of the carbon content of LHS3844b's mantle
effectively reflect that of the bulk composition the planet overall
inherits during formation (Figure~\ref{fig:core}. Therefore, we find
that, in order to produce the C-poor mantle the degassing models
require, and to avoid a magma ocean-derived proto-atmosphere that
would also need to be eroded away in addition to the later degassed
atmosphere, LHS~3844b must form volatile-poor compared to the Earth,
in both single- or multi-stage core formation scenarios.


\subsection{Can Impacts Strip LHS 3844b's Atmosphere?}
\label{sec:impacts}

Infalling material onto a planet either during accretion
\citep{zahnle1990a,zahnle1992a,svetsov2007a,shuvalov2009,schlichting2015},
or through a giant impact
\citep{genda2003b,genda2005,newman1999,schlichting2015} can
potentially remove some or all of an exoplanet's atmosphere. We model
the amount of material needed to strip an atmosphere using the
formalism of \citet{schlichting2015}, which uses analytic self-similar
solutions and full numerical integrations to calculate the fraction of
an atmosphere of scale height, $H$, lost for an individual impactor of
a given radius ($r$) and density ($\rho$), hitting a planet of radius
($R$) with an atmospheric density of $\rho_0$. This formalism yields
three key values: the minimum radius of an impactor able to strip any
atmosphere ($r_\mathrm{min}$), the radius of impactors where it is
considered a giant impact that sends strong shocks through the
atmosphere ($r_{gi}$), and the number ($N$) of impactors of a given
size $r$ needed to entirely strip the atmosphere.

\begin{figure}
  \includegraphics[width=8.5cm]{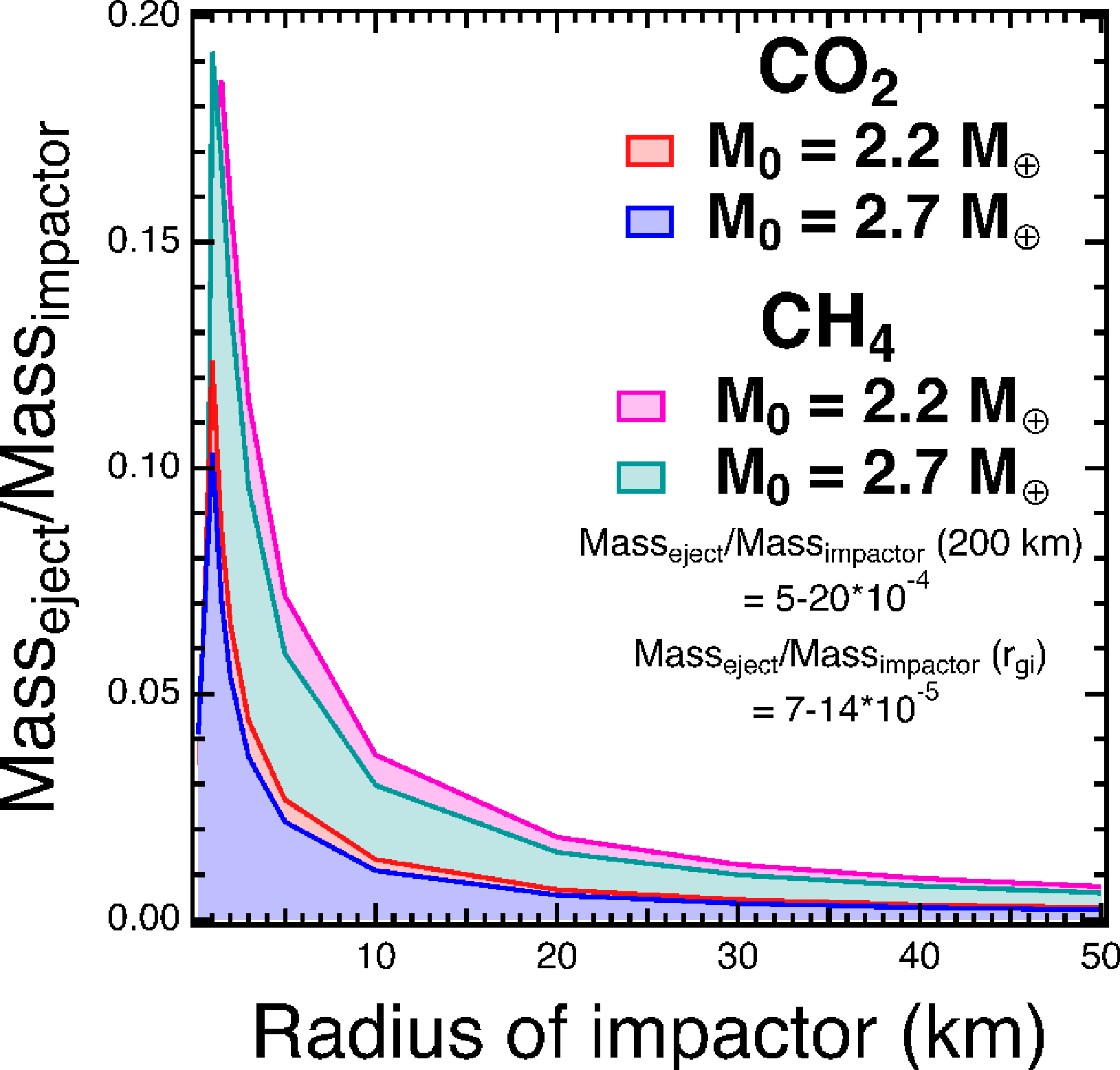}
  \caption{$M_\mathrm{ejected}/M_\mathrm{impactor}$ as a function of
    impactor radius for atmospheres of CO$_2$ ($M_0 = 2.2$~$M_\oplus$,
    red; $M_0 = 2.7$~M$_\oplus$, blue) and CH$_4$ ($M_0 =
    2.2$~$M_\oplus$, pink; $M_0 = 2.7$~$M_\oplus$, teal)).}
  \label{fig:eject}
\end{figure}

We apply this formalism to atmospheres composed of two different
compositions: an oxidized atmosphere containing only CO$_2$ and a
reduced atmosphere containing only CH$_4$. We adopt two accretion
scenarios where LHS~3844b begins at 2.2 and 2.7 $M_\oplus$ and is
accreting the final 0.2~$M_\oplus$ of its mass, rendering our
end-member predicted masses for the planet (Section \ref{bulk}). We
also assume this proto-LHS~3844b is at roughly its current radius of
1.303 $R_\oplus$ as 0.2~$M_\oplus$ of added material will not change
the radius in any significant way \citep{unterborn2019}. This yields
initial gravities for LHS~3844b between 12.7--15.6~m~s$^{-2}$. The
model of \citet{schlichting2015} calculates the minimum radial size of
an impactor to strip any quantity of atmosphere, $r_\mathrm{min}$:
\begin{equation}
  r_\mathrm{min} = H\left(\frac{3\rho_0}{\rho}\right)^{1/3},
\end{equation}
where $H$ is the scale height of the atmosphere, $\rho_0$ is the
density of the atmosphere at the surface, and $\rho$ is the density of
the impactor. We set $\rho$ to 2~g~cm$^{-3}$ below impactor sizes $r =
1000$~km, 3~g~cm$^{-3}$ for $1000 \leq r < 3000$~km and 4~g~cm$^{-3}$
for $3000 \leq r < 5000$~km and 5.5~g~cm$^{-3}$ for impactors of size
$r \geq 5000$~km. To determine scale height, $H$, we assume an
atmospheric temperature of 1000~K and atmospheric pressure of 30 bar
and use the ideal gas law to determine the density of the atmosphere
($\rho_0$) and our gravities derived above. We define $r_{gi}$ as:
\begin{equation}
  r_{gi} = \left(2HR^2\right)^{1/2},
\end{equation}
where $R$ is the radius of the planet (1.303 $R_\oplus$). We calculate
values of $r_{gi}$ between $\sim$1200~km for a pure CO$_2$ atmosphere
and 1700~km for a pure CH$_4$ atmosphere. To quantify the mass of
atmosphere ejected during to an impact relative to the mass of the
impactor for small impact angles,
$M_\mathrm{ejected}/M_\mathrm{impactor}$, we use equation 39 of
\citet{schlichting2015}:
\begin{equation}
  \frac{M_\mathrm{ejected}}{M_\mathrm{impactor}} \backsimeq
  \frac{r_\mathrm{min}}{2r} \left(
  1-\left(\frac{r_\mathrm{min}}{r}\right)^2 \right)
\end{equation}
where $r$ is the radius of the impactor. We find that
$M_\mathrm{ejected}/M_\mathrm{impactor}$ increases with impactor
radius, reaches a maximum at $\sim$1~km, and decreases afterward
(Figure~\ref{fig:eject}). Additionally, CH$_4$ atmospheres are much
easier to strip than CO$_2$ atmospheres are, due to their factor of
$\sim$4 higher scale height. Furthermore, impactors with radii of
$200$~km have $M_\mathrm{ejected}/M_\mathrm{impactor}$ an order of
magnitude higher than impactors meeting the threshold for giant
impacts (Figure~\ref{fig:eject}. While a single giant impact will
strip some fraction of an atmosphere, atmospheric loss by accretion of
many small object is much more efficient process per unit mass,
comparatively \citep{schlichting2015}.

\begin{figure}
  \includegraphics[width=8.5cm]{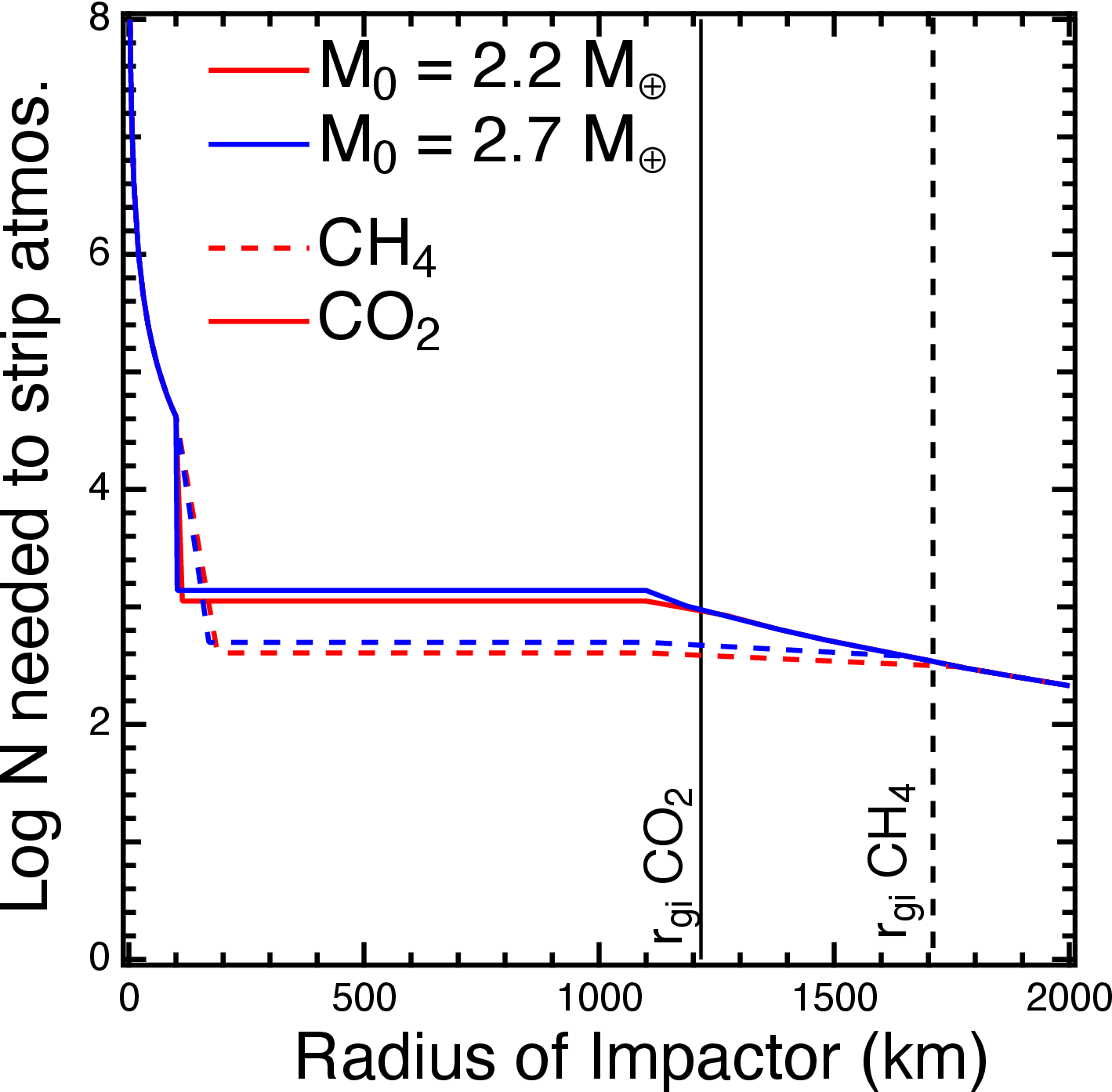}
  \caption{Log number of impactors needed to strip a CO$_2$ (dashed)
    and CH$_4$ atmosphere for initial planet masses of 2.2~$M_\oplus$
    (red) and 2.7~$M_\oplus$ (blue). The radii above which impactors
    are considered giant for atmospheres of CO$_2$ (solid) and CH$_4$
    (dashed) are labeled as black lines.}
  \label{fig:striping}
\end{figure}

The total number ($N$) of impactors of a given size needed to
completely strip an atmosphere is a function of the size of the
atmosphere and the amount of atmosphere ejected due to an impact of a
given size \citep{schlichting2015}. To simplify this model, we assume
that these impactors are not carrying additional volatiles to be added
to the atmosphere. Using equations 47, 49, and 51 of
\citet{schlichting2015}, which assume the impactor is traveling at
roughly the escape velocity of the planet, we find that, for a CO$_2$
atmosphere, roughly $N \sim 10^{3-8}$ small impactors ($r < r_{gi}$)
are needed to entirely strip the atmosphere, with the exact value
depending on the impactor's size (Figure \ref{fig:striping}, top). We
find slightly smaller values of $N$ for CH$_4$ atmospheres, likely due
to the larger scale height compared to CO$_2$. Above $r_{gi}$ for both
species, $N$ slowly drops as impactor radius increases, reaching
$\sim$30 accretionary impacts needed for Mars-sized impactors and
$\sim$3 needed for Earth-sized impactors.

\begin{figure}
  \includegraphics[width=8.5cm]{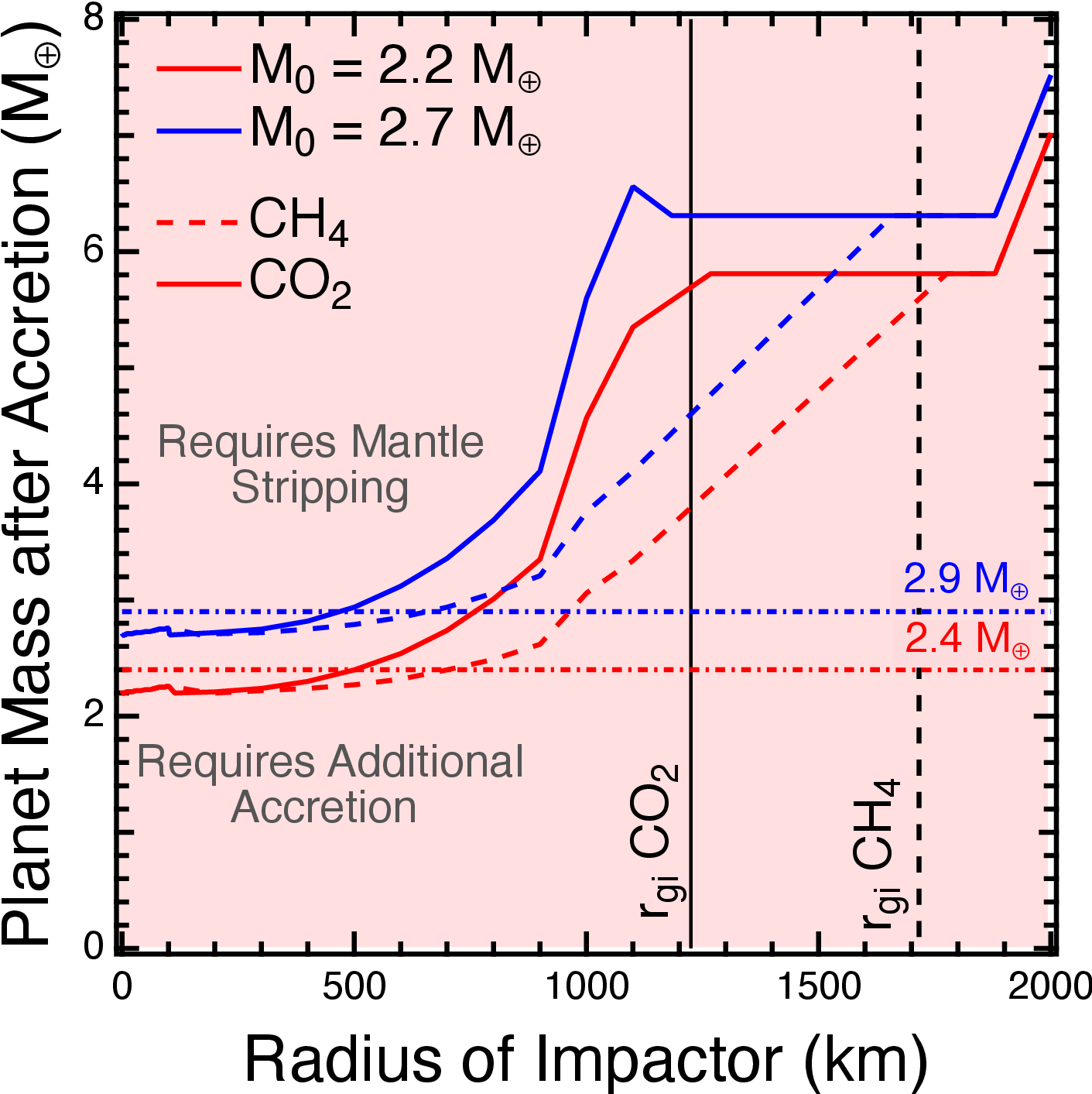}
  \caption{Final mass of LHS~3844b starting at 2.2~$M_\oplus$ (red)
    and 2.7~$M_\oplus$ (blue) after accreting the $N$ impactors of a
    given radius needed to strip its CO$_2$ (solid) or CH$_4$ (dashed)
    atmosphere. Our final estimates for the mass of LHS 3844b if it
    had a relatively small core (2.4~$M_\oplus$) or relatively large
    core (2.9~$M_\oplus$) are shown as red and blue dot-dashed lines,
    respectively. Radii above which impactors are considered giant for
    atmospheres of CO$_2$ (solid) and CH$_4$ (dashed) are labeled as
    black lines.}
  \label{fig:totmass}
\end{figure}

While accretion can strip an atmosphere, it also adds to the overall
mass of a planet. Our initial masses require only 0.2 $M_\oplus$ of
material to reach our final estimated masses of LHS~3844b (2.4 and 2.9
$M_\oplus$). For those impactors where the total amount of accreted
material exceeds 0.2~$M_\oplus$, some fraction of the protoplanet must
also be stripped to space upon impact. For protoplanets that have
undergone core segregation, the mantle will be stripped, leaving
behind a planet with a relatively large core mass fraction. For a
CO$_2$ atmosphere, the size of impactor where mantle stripping becomes
necessary is $\sim$450~km for both our small and large initial planet
masses (Figure~\ref{fig:totmass}, solid). For a CH$_4$ atmosphere,
these values rise to $\sim$900~km (Figure~\ref{fig:totmass},
dashed). These differences due to atmospheric composition are purely a
function of the factor of $\sim$4 difference in the scale heights of
these gasses. Many more than $N$ impactors of these sizes can accrete;
however, mantle stripping would become required in such a case, in
order to reduce the overall mass. This does not mean that impactors
below this size cannot strip mantle, but rather that, if all infalling
material is above this size, mantle stripping \textit{must} occur in
order to match our mass estimates of LHS~3844b. For all of our models,
giant impact scenarios require significant mantle stripping to match
LHS~3844b's mass regardless of atmosphere composition.

These previous models assume that the impactors velocity is roughly
that of the escape velocity. However, as the velocity of impact
changes, so does the relative amount of atmosphere ejected. For
example, both Earth and LHS~3844b require $\sim$30 impactors of Mars
size ($\sim$3400~km) to strip their atmospheres completely when
traveling at the escape velocity ($v_\mathrm{escape}$). The current
orbital velocity of LHS 3844b is $\sim$150~km/s. Using equation~32
from \citet{schlichting2015}, for a 500 km object to remove a 1/$N$
amount of atmosphere ($N = 300$) from a 2~$M_\oplus$ LHS~3844b, it
must travel at $\sim$70 times the escape velocity for both a pure
CO$_2$ and CH$_4$ atmosphere. For a Mars-sized object ($r = 3400$~km,
$N = 30$), this velocity is reduced to essentially the escape
velocity. If accreting material impacts at speeds slower than these,
$N$ must increase in order to fully strip the atmosphere.

\begin{figure}
  \includegraphics[width=8.5cm]{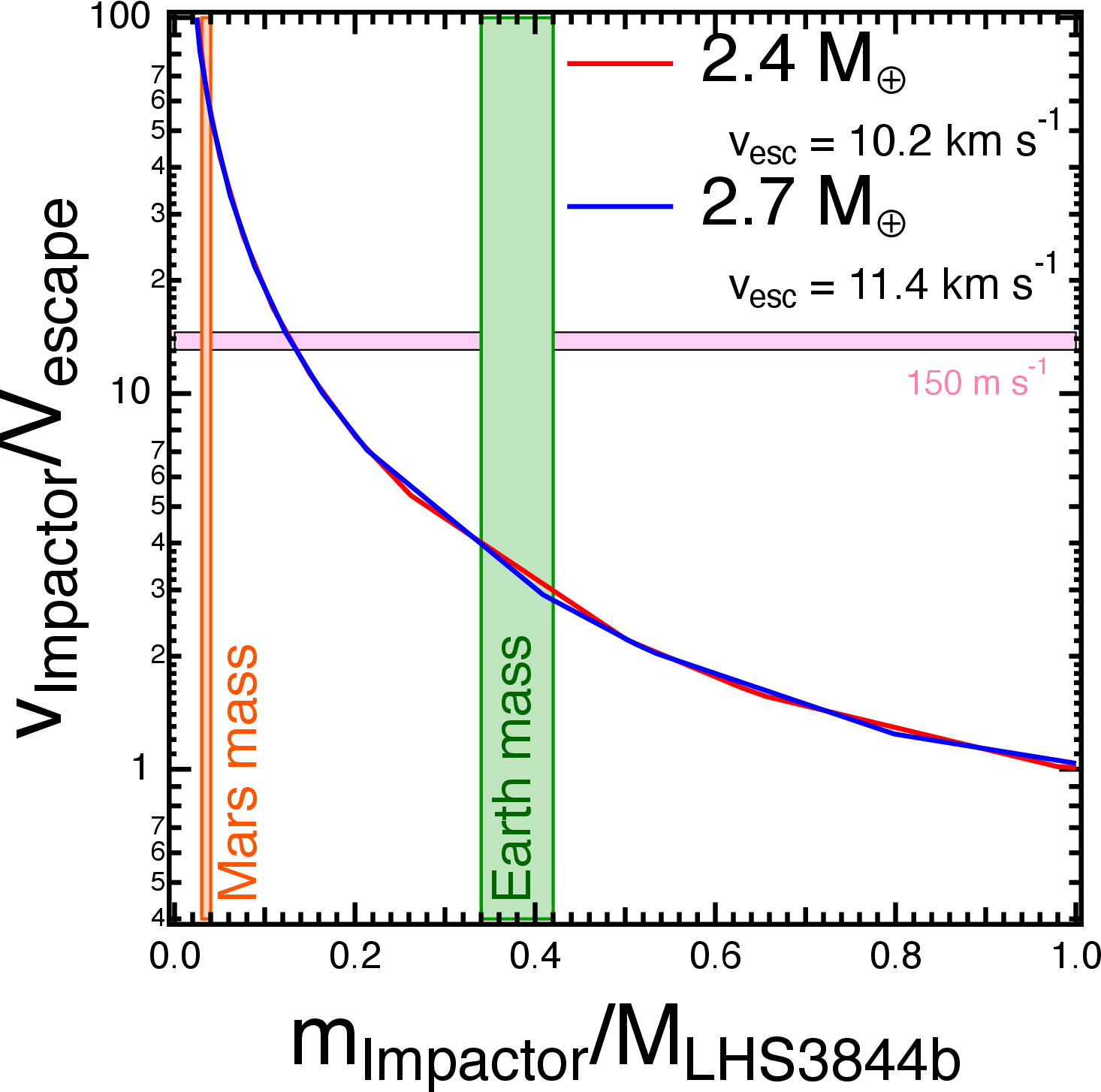}
  \caption{Relative velocity of impactor to the escape velocity of LHS
    3844b needed to entirely strip an atmosphere with a single impact
    as a function of the mass ratio of the impactor to the proto-LHS
    3844b for initial planet masses of 1~$M_\oplus$ (blue),
    2~$M_\oplus$ (red), and 2.7~$M_\oplus$ (black). Ranges of relative
    masses for Mars (orange) and Earth (green) across our estimated
    masses of LHS 3844b are shown for reference.}
  \label{fig:vesc}
\end{figure}

While $N$ is a measure of how many accreting impacts traveling at
roughly $v_\mathrm{escape}$ are needed to strip an atmosphere, we can
also estimate the impact velocity ($v_\mathrm{impactor}$) needed to
entirely strip the atmosphere in a single giant impact after LHS~3844b
has fully formed with a radius of 1.303~$R_\oplus$ and between
2.4--2.9~$M_\oplus$. Using equation~32 of \citet{schlichting2015} for
an isothermal atmosphere and setting the fraction of atmosphere loss
to 1, we estimate the relative velocity of an impactor to the planet's
escape velocity ($v_\mathrm{impactor}/v_\mathrm{esc}$) needed to
completely strip the atmosphere in one impact as function of the
impactor's mass relative to LHS 3844b's
($m_\mathrm{impactor}/M_\mathrm{LHS 3844b}$). We find that, as
$m_\mathrm{impactor}/M_\mathrm{LHS 3844b}$, the
$v_\mathrm{impactor}/v_\mathrm{escape}$ needed to eject an atmosphere
entirely decreases (Figure~\ref{fig:vesc}). The value of
$v_\mathrm{escape}$ varies with the initial mass of the planet,
between 10.2 and 11.4~km/s for our range of initial mass models. For a
Mars-sized object, it must travel $\sim$100 times the escape velocity,
whereas if this Mars-sized object was to hit the Earth, it would be
required to travel at only $\sim$10 times the escape velocity
($m_\mathrm{impactor}/m_\oplus = 0.1$). For an Earth-sized impactor
hitting LHS~3844b, it need only travel at 1.3--1.4 times the escape
velocity to strip the atmosphere entirely. These velocities are
roughly 70 times the orbital velocity for a Mars-size object and only
10\% for an Earth-size object. We calculate the minimum size of an
object that can strip an atmosphere entirely in one impact while
traveling at the escape velocity to be 11--12\% of the mass of LHS
3844b, or 0.26--0.35~$M_\oplus$. Numerical results have found that
multiple giant impacts like these are likely in models of planets
Earth-size and slightly larger ($M < 1.6$~$M_\oplus$) in the Solar
protoplanetary disk \citep{quintana2016}, with Kepler-107c showing
evidence that it underwent a mantle-stripping giant impact
\citep{bonomo2019}. \citet{quintana2016} also found, however, that the
planets in their model only experience an average of three giant
impacts (up to a maximum of eight in their models), all of which were
unlikely to fully strip its atmosphere or oceans. Whether this is also
true for systems like LHS~3844b is not known. However, if the planet
was impacted by a single Earth-size impactor traveling at or below the
orbital velocity of LHS~3844b, complete atmospheric ejection and
significant mantle stripping is likely. Comparatively, a single impact
from a Mars-sized object is a less plausible scenario for stripping
the atmosphere of LHS~3844b due to the high impact velocity needed.

Either through the accretion of small objects ($r < r_{gi}$) or single
giant impact events, atmospheric stripping is only effective if there
is an atmosphere to eject. Degassing from a planet's interior is not
an instantaneous process. If this final 0.2~$M_\oplus$ of mass
accretes onto LHS 3844b prior to degassing from the magma ocean or
subsequent degassing from the solid mantle, the material will simply
accrete as normal and no atmospheric stripping will
occur. \citet{eilkinstanton2011b} found that significant volatile
degassing due to the solidification of the magma ocean happens over 5
Myr. Pebble accretion tends to form planets via accretion relatively
rapidly, on the order of $<1$ Myr \citep{johansen2017}, making it
unlikely that accretionary impacts during the magma ocean phase will
have any atmosphere to strip. Unless material is delivered later,
after the magma ocean phase, will impacts by smaller objects be a
viable scenario for stripping LHS 3844b's atmosphere. However, in the
case of impact stripping after the magma ocean stage, accreting
material would have to arrive $> \sim 100$ Myrs after magma ocean
solidification for any significant atmosphere to exist, based on our
mantle outgassing models. Something akin to a late heavy bombardment
would be required in order to strip the outgassed atmosphere after
magma ocean solidification, either with smaller impactors or giant
impactors. As impactors with mass above 0.26--0.35~$M_\oplus$ need
only travel at or below the orbital velocity of LHS 3844b to entirely
strip the atmosphere, giant impactors can not be ruled out as an
explanation for LHS 3844b's lack of atmosphere. In the case of these
giant impacts, however, significant mantle stripping is likely. As
such, if upon measurement of LHS 3844b's mass and host-star
composition, if LHS~3844b should have a density higher than
characteristic of a nominally rocky planet with the host star being
comparatively Fe-poor, this would imply that the planet underwent a
significantly massive, mantle-stripping, giant impact event, akin to
those expected for observed super-Mercuries \citep{bonomo2019}.

\section{Discussion}
\label{discussion}

Our models pose three general possibilities for the composition and
dynamical history of LHS~3844b: 1) a giant impact ejected all of its
atmosphere; 2) it did not undergo degassing at all, due to its
geophysical state; and/or 3) it formed C-poor relative to the Earth
(Figure~\ref{fig:all_hists}, panels A, F, K). Interestingly, for
planets where volcanism and degassing are suppressed entirely, planets
larger than Earth core sizes also favor low-C abundances, with a
slight favoring of low C for Earth-like core sizes
(Figure~\ref{fig:no_outgas}). As discussed in
Section~\ref{sec:impacts}, we cannot definitively rule out a
LHS~3844b's atmosphere being ejected by a giant impact, but note that
this scenario will likely leave the planet as an Fe-enriched "super
Mercury." If LHS~3844b is found to have a high density upon
measurement of its mass, this will lend credence to this scenario
explaining the planet's lack of atmosphere.

Scenario 2 requires that a planet forms either cold with an initial
mantle potential temperature $<1800$~K, have an interior composed of
high-viscosity minerals, or that it forms with a radiogenic heat
budget between 50--60\% of the Earth's. Most likely, a planet would
have to have all three of these aspects to prevent melting. Any planet
that formed cold with high viscosity, but with an Earth-like
radiogenic heat budget, would simply heat up as energy is released
into the mantle from radioactive decay, raising the potential
temperature and triggering melting. Similar arguments can be made for
each of the other combinations. A test for scenario 1 would be to
measure the radionuclide abundances of the host-star LHS 3844, which
provide observational constraints for the radiogenic heat budget of
the planet \citep{unterborn2015}.

Scenario 3 argues that sub-Earth concentrations of C are the most
likely explanation for the present thin, $<10$ bar atmosphere. This is
due to the low partition coefficient of C limiting how much can be
stored in the planet's core, thus producing a mantle with a C
concentration nearly that of the bulk planet's. For bulk planet
compositions at or above the Earth's, these mantle concentrations are
well above the smallest that we predict will produce a $<10$ bar
atmosphere today (0.002 wt\%;
Figure~\ref{fig:activityrotation}). Therefore, despite the relatively
high age of the host star (6--9~Gyr), the atmospheric stripping rate
would have to be at least a factor of 10 larger than the estimated
upper bound for an Earth-like C abundance to result in a present-day
atmosphere $<$ 10 bar (see Equation~\ref{eq:c_tot_max}).

A relatively volatile-poor composition of LHS~3844b suggests the
planet formed interior to its water snow-line (or ice-line). Had
LHS~3844b formed beyond the snow-line, it would likely contain
significant amounts of water, similar to those inferred for TRAPPIST-1
\citep{unterborn2018a,unterborn2018b,grimm2018}. This water would then
be outgassed at the same time as CO$_2$, contributing even more mass
to the atmosphere, as liquid water is not stable on LHS 3844b's
surface. Assuming LHS~3844b did form outside the snow-line, a
conservative estimate of its primordial water content would be
$\sim$1\% by mass. For the mass range described in Section~\ref{bulk},
this equates to $1.4-1.7\times10^{23}$~kg of water. If the mass-loss
rate due to atmospheric erosion is the same as our highest assumed
value for CO$_2$ (300~kg~s$^{-1}$), only $8.9\times10^{19}$ kg of
water could be removed from the planet. Therefore, if LHS~3844b had
formed beyond the snow-line, a thick H$_2$O (or possibly O$_2$, due to
hydrogen escape) atmosphere would still remain, unless the escape rate
was $\sim$4 orders of magnitude larger than our
estimates. Interestingly, scenario 2 also implies a relatively
water-poor composition of LHS~3844b. This is due to the fact that, as
water is added to mantle rocks, it lowers both their melting
temperature \citep{hirschmann2006b} and viscosity
\citep{karato2011a}. This allows for melting and degassing to occur at
lower temperatures and shallower depths than those of the dry mantle
case modeled here. C-rich ices condense at much lower temperatures
(and greater orbital distances) than water in disks
\citep{lodders2003}. Any planet forming further out in the disk where
C-rich ices are stable would also form in a region where condensed
water is also stable. A water- or carbon-rich origin of LHS~3844b is
unlikely in both scenarios 2 and 3.

\begin{figure}
  \includegraphics[width=8.5cm]{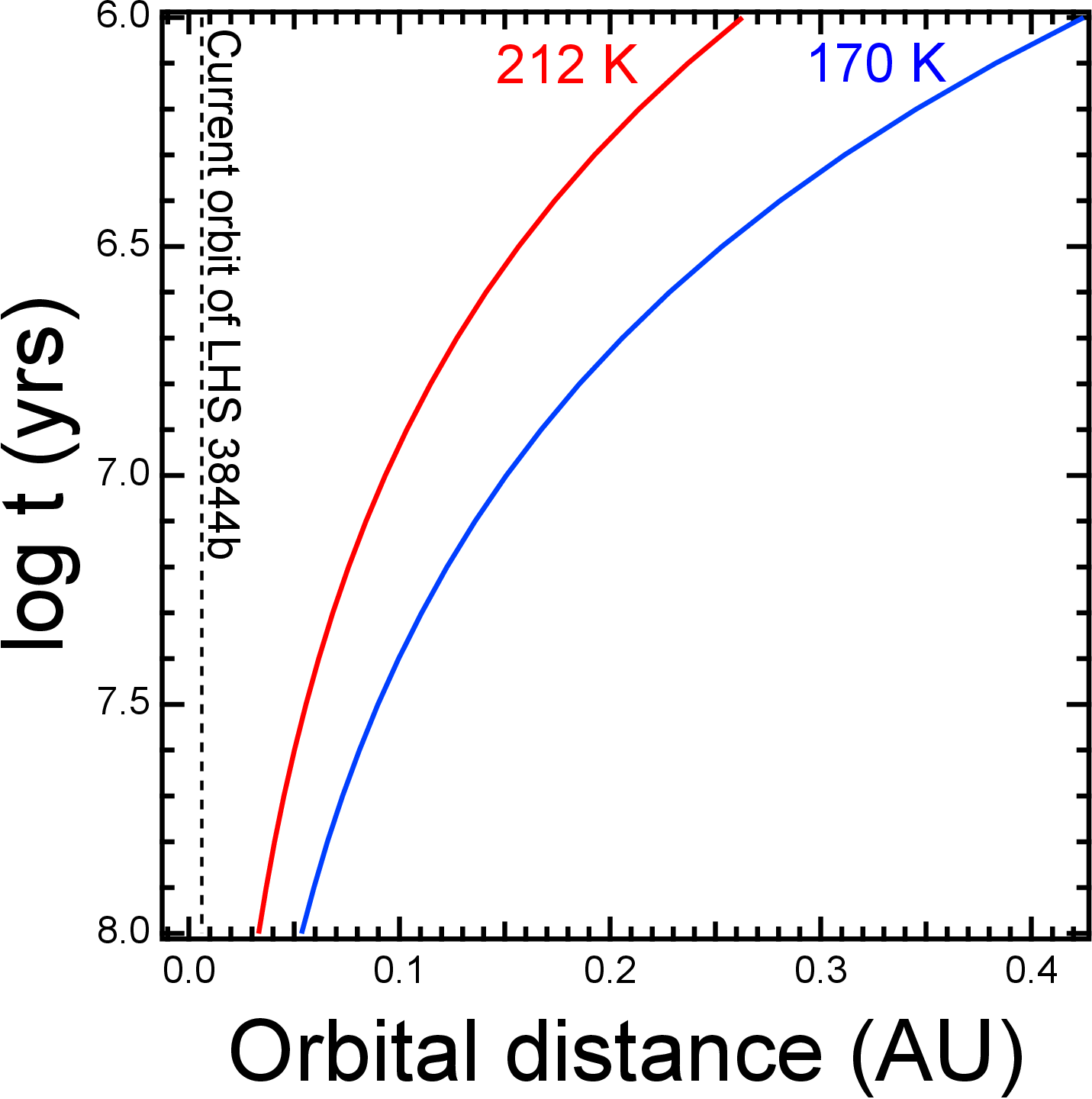}
  \caption{Calculated orbital distance of the water snow-line for
    LHS~3844b as a function of time for water condensation
    temperatures of an M-dwarf disk (212~K; red) and the solar nebula
    (170~K; blue). The current orbital distance of LHS~3844b is shown
    as a black dashed line.}
  \label{fig:snowlines}
\end{figure}

If LHS~3844b formed interior to the water snow-line, this allows us to
constrain the degree to which it migrated as it
formed. \citet{unterborn2018a} calculated the location of the
snow-line in M-dwarf disks, assuming a passively heated, flared disk
\citep{chiang1997}. From \citet{unterborn2018a}, we estimate the
location of the snow-line to be located at 0.26~AU at 1 Myr
(Figure~\ref{fig:snowlines}, assuming the reported stellar mass of
0.151~$M_\odot$ and estimating luminosity of LHS~3844 using the
stellar evolution models of
\citet{baraffe2002}. \citet{unterborn2018a} calculated the
temperature of water condensation as 212~K, which is greater than that
of the solar nebula (170~K) owing to the greater surface density of M
dwarf disks compared with the solar nebula.

The location of the water snow-line thus sets the maximum orbital
distance at which LHS~3844b could have formed. If LHS~3844b formed
later than 1~Myr, this maximum distance decreases as the disk cools
(Figure~\ref{fig:snowlines}). As LHS~3844b cannot cross exterior to
the snow-line, this also sets the rate at which it must migrate. If
LHS~3844b formed early (1~Myr), a rapid migration rate followed by
slower one is possible in order to remain interior to the
snow-line. If LHS~3844b formed at 10~Myr, a much slower migration rate
and a smaller migration distance are possible
\citep{lykawka2013,izidoro2014b}. It is also possible that LHS~3844b
formed entirely in place with no migration. This connection between
observed composition and disk chemistry clearly provides constraints
on the formation dynamics of the planet itself and can be applied to
other systems.

The volatile-poor nature of LHS~3844b provides additional clues to its
formation history. \citet{kreidberg2019a} place an upper limit for
LHS~3844b's surface water content of $<0.02$ wt\%. This is very
similar to the Earth's concentration of $<0.01$ wt\% water, which is
partitioned between the surface and mantle. These water mass fractions
are exceedingly dry compared to even the driest chondritic meteorites
\citep[CV: 2.5 wt\%, CO: 0.63 wt\%,][]{wasson1988,mottl2007b}. Similar
depletion is seen for carbon in the case of the Earth
\citep{mcdonough1995}. All of this is despite water ice being stable
at 1 AU while the disk was present
\citep{oka2011}. \citet{morbidelli2016a} argues that the inner Solar
System is particularly dry due to the presence of proto-Jupiter
preventing the drift of volatile-rich material inward from beyond the
snow-line. Looking to the TRAPPIST-1 system, we see that TRAPPIST-1b
and c both likely formed interior to the primordial snow-line during
the disk lifetime, yet are inferred to contain significant water
fractions \citep{unterborn2018a}. This suggests that volatile-rich
material can accrete onto interior planets from beyond the snow line
in M-dwarf systems, yet this process seems to not have occurred for
LHS 3844b. The presence of a larger, undetected planet orbiting LHS
3844 that similarly restricted volatile-rich material to LHS 3844b may
then explain our results. For example, consider a planet located at
the snow-line for the $T = 212$~K youngest age scenario shown in
Figure~\ref{fig:snowlines} (2.6~AU). Planets with masses in the range
of Neptune to Jupiter located at that snow-line would induce radial
velocity signals with semi-amplitudes in the range 8--143~m\,s$^{-1}$
with a period of $\sim$125~days. The radial velocities obtained by
\citet{vanderspek2019} were insufficient to place constraints on the
presence of such additional planetary companions in the system, and
would require higher cadence monitoring over an extended period with
precision capable of detecting $\sim$10~m\,s$^{-1}$ signals. Whether a
potentially undetected planet need be Jupiter-mass to create a similar
pressure gradient to block inflow from beyond the ice line is thus
presently unknown.

Our model excludes magnetic fields, which are typically assumed to
protect a planet from atmospheric erosion and therefore effectively
lower the erosion rate. The interaction between the stellar wind and a
planet's secondary atmosphere is complex, but has been explored for
Earth-size planets with both Earth and Venus-like compositions
\citep[e.g., ][]{dong2017a,dong2018a,dong2020}, potentially preserving
Venus-like atmospheres on super-Earths for $>10$~Gyr. Recent studies,
however, have shown that magnetic fields may not help retain an
atmosphere at all \citep{gunell2018a}. Whether the presence of a
magnetic field would lower the atmospheric erosion rate for LHS 3844b
is uncertain. We note, however, that if magnetic fields do
substantially lower atmospheric stripping rates, including this effect
for LHS 3844b in our models would lower our estimated maximum C budget
that is able to produce a $<10$ bar atmosphere, by decreasing the
atmospheric loss rate, $E$, in Equation~\ref{eq:c_tot_max}. That is, a
protective magnetic field would act to favor an even more C depleted
planet than we estimate here in order to satisfy the observational
constraints.

Our models also consider only primordial heat and radioactive heat
production for LHS~3844b. However, additional heat sources could be
important. Electromagnetic induction heat is potentially significant
for planets on close-in orbits \citep{kislyakova2017,kislyakova2018},
and tidal heating could be significant if LHS 3844b has an eccentric
orbit \citep[e.g.][]{driscoll2015}. As \citet{kreidberg2019b}
indicates that the surface of LHS~3844b is solid, any additional heat
sources beyond radioactive heat-producing elements must be small
enough that they do not result in a fully molten, magma ocean
state. The addition of tidal and magnetic induction heat sources at a
moderate level, which still result in a mostly solid mantle, would
only act to reinforce our main findings. Additional heat sources would
hasten the already rapid outgassing of volatiles from the interior to
the atmosphere, thus further highlighting that LHS~3844b will likely
have outgassed its interior volatile supply to the surface. Moreover,
additional heat sources would make the already unlikely case of LHS
3844b having too high a mantle viscosity and too little internal heat
to experience significant outgassing even less likely. As a result,
whatever volatiles the planet acquired are likely to have been
outgassed to the atmosphere. In order for LHS~3844b to have a thin
atmosphere today, the volatile abundance must have been low, the rate
of atmospheric loss must have been at least an order of magnitude
higher than our estimated upper limit based on Proxima Cen b, or
atmospheric stripping by a giant impact must have taken place.

LHS~3844b is considered an ultra-short period (USP) planet, and we
placed this system in context with other USP planets by comparing it
to 55~Cancri (55~Cnc) and 55~Cnc~e. We focus on the differences that
could cause LHS~3844b to have no atmosphere, while studies have shown
that 55~Cnc~e has an atmosphere \citep{bourrier2018c}. The mass for
55~Cnc determined with interferometric and photometric observations by
\citet{ligi2016} is $0.85\pm0.24$~$M_\odot$, and a rotation period of
$38.8\pm0.5$ days was determined by \citet{bourrier2018c}. Together,
these give a Rossby number of $R_o = 2.2 \pm 0.9$; the large
uncertainty is due to the large uncertainty of the stellar
mass. Making use of the mass derived by \citet{ligi2016}, which is
based upon isochrones ($0.874\pm0.013$~$M_\odot$), provides stronger
constraints on $R_o = 2.2 \pm 0.06$. The Rossby number of 55~Cnc is
extremely likely to be significantly higher than that of LHS~3844,
meaning that 55~Cnc is a less active star. Additionally,
\citet{bourrier2018c} provide a semi-major axis for 55~Cnc~e of
$0.01544\pm0.00005$~AU, but \citet{vanderspek2019} give a semi-major
axis of $0.00622\pm0.00017$~AU for LHS~3844b. The differences in
stellar properties and semi-major axes between these two planets may
result in a measurable atmosphere for 55~Cnc~e despite a low volatile
budget, although detailed observational analysis indicates that
55~Cnc~e likely has a magma surface \citep{demory2016b}.

\section{Conclusions}
\label{conclusions}

The degassing of a planet during the first $\sim 100$s of Myrs and the
subsequent erosion of the atmosphere due to stellar activity are
complicated processes with numerous combinations resulting in a
variety of outcomes. High stellar activity with low degassing rates
can lead to a prevalence of thin secondary atmospheres, while low
stellar activity with high degassing rates can result in a retained
combination of primary and secondary atmospheres. Investigating the
relative contribution of these key processes to atmospheric
sustainability and evolution is thus critical for understanding the
origin of observed atmospheric masses and abundances. LHS~3844b is a
fascinating case in which the atmospheric evolution and influence of
stellar activity may be explored in detail. Our age estimate for the
host star of $\sim$7.8~Gyrs provides sufficient time for significant
atmospheric loss to have occurred for the planet.

However, whether atmospheric loss over the course of $\sim$6--9~Gyrs
can remove enough mass to explain the currently observed thin
atmosphere of LHS~3844b depends on the geological characteristics of
the planet. Specifically, our geophysical models of the planet's
thermal and volcanic history indicate that, with a mantle volatile
budget similar to Earth's, $\sim$200 bar of CO$_2$ would be
outgassed. Such a large atmosphere cannot be removed by even the
fastest stripping rate we consider. We find that a giant impact of a
0.26--0.35~$M_\oplus$ object traveling at the orbital velocity of
LHS~3844b could entirely eject this atmosphere, but it would also
strip considerable amounts of mantle, resulting in LHS 3844b being an
Fe-rich "super-Mercury." In the absence of a giant impact, LHS~3844b
was therefore most likely volatile poor compared to Earth, due to the
low partitioning of C into the core of LHS 3844b under both oxidizing
and reducing formation conditions. The minimum mantle C concentration
we find that is able to produce a $<10$ bar atmosphere today is
$\sim$10 times lower than estimates for the Earth. Volatile
inventories of this size produce atmospheres of $\approx$20--25~bar,
which can be stripped to $< 10$ bar over the course of 6--9~Gyrs. It
is also possible to suppress volcanism and outgassing entirely if
LHS~3844b has a high overall mantle viscosity ($\sim$100 times more
viscous than Earth's) and low abundance of radioactive heat-producing
elements ($\sim$50\% that of Earth's). However, this unique
combination of parameters that can suppress outgassing was rarely seen
in our models and also implies a volatile-poor composition for
LHS~3844b. These results are consistent with the recent findings of
\citet{kite2020c}, who argued that planets forming close to M-dwarf
stars with high planet equilibrium temperatures must have a higher
volatile budget than the Earth in order to retain an atmosphere today.

These results imply that LHS~3844b formed both water and C-poor
compared to the Earth. We propose then that it formed interior the the
snow-line, which provides an upper limit on the degree to which it
could have migrated. Furthermore, should planets in M-dwarf systems
form via pebble accretion, as in our Solar System, these results
suggest the presence of a more massive companion at a larger orbital
distance. This companion may have restricted the inflow of
volatile-rich material to the inner disk where LHS~3844b formed, in a
manner similar to Jupiter's role in restricting volatile-rich material
to the Earth. Early migration and other dynamical changes to the orbit
may have resulted in significant modifications to the radiation
environment of the planet and subsequent atmospheric loss rates,
though these are expected to have occurred early enough to not have
played a major role in the overall atmospheric evolution.

The results presented here have been applied to a single case that
represents one of the very significant limits on the atmospheric
sustainability of a super-Earth planet in close proximity to an M
dwarf. With the expectation of numerous further terrestrial planet
discoveries around bright stars by {\it TESS}
\citep{sullivan2015,barclay2018}, there will be additional
opportunities to study the relationship between stellar properties and
atmospheric evolution. Our results point to the role geoscience can
play in contextualizing astrophysical observations, providing key
constraints for astrophysical models, and predicting observational
tests. As we move toward deeper characterization of individual
exoplanetary systems with little directly observed data, these
interdisciplinary interactions become even more vital.


\section*{Acknowledgements}

The authors would like to thank the anonymous referees, whose feedback
improved the quality of the paper. R.M.R. acknowledges support from
the YCAA Prize Postdoctoral Fellowship. This research has made use of
the following archives: the Habitable Zone Gallery at hzgallery.org
and the NASA Exoplanet Archive, which is operated by the California
Institute of Technology, under contract with the National Aeronautics
and Space Administration under the Exoplanet Exploration Program. The
results reported herein benefited from collaborations and/or
information exchange within NASA's Nexus for Exoplanet System Science
(NExSS) research coordination network sponsored by NASA's Science
Mission Directorate.




\end{document}